\pdfoutput=1
\documentclass[11pt,twoside,a4paper,cmspaper,final,collab]{cms-tdr}
\def\svnVersion{4be76fa-D}\def\svnDate{2021/03/12}\def\cmsCernNoTag{CERN-EP-2020-211}\def\cmsCernDate{\today}\def\cmsMessage{"Published in the Journal of High Energy Physics as \href{http://dx.doi.org/10.1007/JHEP03(2021)095}{\doi{10.1007/JHEP03(2021)095}.}"}
\begin{document}\cmsNoteHeader{TOP-19-001}

\newcommand{\quotes}[1]{``#1''}

\renewcommand{\lumi}{\ensuremath{41.5\fbinv}\xspace}

\newcommand{\PAlepton}{\ensuremath{\overline{\ell}}}
\newcommand{\Pggx}{\ensuremath{\PGg^{*}}\xspace}
\newcommand{\ttH}{\ensuremath{ \PQt\PAQt\PH}\xspace}
\newcommand{\ttZ}{\ensuremath{ \PQt\PAQt\cPZ}\xspace}
\newcommand{\ttW}{\ensuremath{ \PQt\PAQt\PW}\xspace}
\newcommand{\ttll}{\ensuremath{ \PQt\PAQt \Pl \PAl}\xspace}
\newcommand{\ttlnu}{\ensuremath{ \PQt \PAQt \Pl \PGn}\xspace}
\newcommand{\ttgamma}{\ensuremath{ \PQt \PAQt \PGg}\xspace}
\newcommand{\tllq}{\ensuremath{ \PQt \Pl \PAl \cPq}\xspace}
\newcommand{\tHq}{\ensuremath{ \PQt\PH\cPq}\xspace}

\newcommand{\tW}{\ensuremath{ \PQt\PW}\xspace}

\newcommand{\tZq}{\ensuremath{ \PQt \cPZ \cPq}\xspace}

\newcommand{\WJets}{\ensuremath{\PW\text{+jets}}\xspace}
\newcommand{\ZJets}{\ensuremath{\cPZ\text{+jets}}\xspace}
\newcommand{\ttbarJets}{\ensuremath{\ttbar\text{+jets}}\xspace}

\renewcommand{\ss}{\ensuremath{\text{ss}}}

\newcommand{\hc}[1]{{}^\ddagger #1}
\newcommand*{\tmp}[4]{\ensuremath{
    {#4
    \ifx\empty#3\empty\ifx\empty#1\empty\else^{#1}\fi\else^{#1(#3)}\fi
    \ifx\empty#2\empty\else_{#2}\fi}
}}
\newcommand*{\qq }[4][]{\tmp{#2}{#3}{#4}{#1{O}}}

\newcommand{\FDF}{(\varphi^\dagger i\!\!\overleftrightarrow{D}_\mu\varphi)}
\newcommand{\FDFI}{(\varphi^\dagger i\!\!\overleftrightarrow{D}^I_\mu\varphi)}

\newcommand{\sW}{\ensuremath{s_{\mathrm{W}}}\xspace}
\newcommand{\cW}{\ensuremath{c_{\mathrm{W}}}\xspace}

\newcommand{\muR}{\ensuremath{\mu_{\mathrm{R}}}}
\newcommand{\muF}{\ensuremath{\mu_{\mathrm{F}}}}
\newcommand{\muRF}{\ensuremath{\mu_{\mathrm{R,F}}}}

\newcommand*{\eftOp}[4]{\ensuremath{
    {#4
    \ifx\empty#3\empty\ifx\empty#1\empty\else^{#1}\fi\else^{#1(#3)}\fi
    \ifx\empty#2\empty\else_{#2}\fi}
}}
\newcommand{\ctp}  {\eftOp{}{ \PQt\varphi}{}{c}\xspace}
\newcommand{\ctpI} {\eftOp{I}{ \PQt\varphi}{}{c}\xspace}
\newcommand{\cpQM} {\eftOp{-}{\varphi Q}{}{c}\xspace}
\newcommand{\cpQa} {\eftOp{3}{\varphi Q}{}{c}\xspace}
\newcommand{\cpt}  {\eftOp{}{\varphi \PQt}{}{c}\xspace}
\newcommand{\cptb} {\eftOp{}{\varphi \PQt \PQb}{}{c}\xspace}
\newcommand{\cptbI}{\eftOp{I}{\varphi \PQt \PQb}{}{c}\xspace}
\newcommand{\ctW}  {\eftOp{}{ \PQt\PW}{}{c}\xspace}
\newcommand{\ctWI} {\eftOp{I}{ \PQt\PW}{}{c}\xspace}
\newcommand{\ctZ}  {\eftOp{}{ \PQt\cPZ}{}{c}\xspace}
\newcommand{\ctZI} {\eftOp{I}{ \PQt\cPZ}{}{c}\xspace}
\newcommand{\cbW}  {\eftOp{}{ \PQb\PW}{}{c}\xspace}
\newcommand{\cbWI} {\eftOp{I}{ \PQb\PW}{}{c}\xspace}
\newcommand{\ctG}  {\eftOp{}{ \PQt G}{}{c}\xspace}
\newcommand{\ctGI} {\eftOp{I}{ \PQt G}{}{c}\xspace}

\newcommand{\cQla} {\eftOp{3}{Q\ell}{\ell}{c}\xspace}
\newcommand{\cQlM} {\eftOp{-}{Q\ell}{\ell}{c}\xspace}
\newcommand{\cQe}  {\eftOp{}{Q\Pe}{\ell}{c}\xspace}
\newcommand{\ctl}  {\eftOp{}{ \PQt \ell}{\ell}{c}\xspace}
\newcommand{\cte}  {\eftOp{}{ \PQt \Pe}{\ell}{c}\xspace}
\newcommand{\ctlS} {\eftOp{S}{ \PQt}{\ell}{c}\xspace}
\newcommand{\ctlSI}{\eftOp{SI}{ \PQt}{\ell}{c}\xspace}
\newcommand{\ctlT} {\eftOp{T}{ \PQt}{\ell}{c}\xspace}
\newcommand{\ctlTI}{\eftOp{TI}{ \PQt}{\ell}{c}\xspace}

\newlength\cmsTabSkip\setlength{\cmsTabSkip}{1ex}

\cmsNoteHeader{TOP-19-001}
\title{Search for new physics in top quark production with additional leptons in proton-proton collisions at \texorpdfstring{$\sqrt{s} = 13\TeV$}{sqrt(s) = 13 TeV} using effective field theory}

\date{\today}

\abstract{
Events containing one or more top quarks produced with additional prompt leptons are used to search for new physics within the framework of an effective field theory (EFT). The data correspond to an integrated luminosity of \lumi of proton-proton collisions at a center-of-mass energy of 13\TeV at the LHC, collected by the CMS experiment in 2017. The selected events are required to have either two leptons with the same charge or more than two leptons; jets, including identified bottom quark jets, are also required, and the selected events are divided into categories based on the multiplicities of these objects.  Sixteen dimension-six operators that can affect processes involving top quarks produced with additional charged leptons are considered in this analysis. Constructed to target EFT effects directly, the analysis applies a novel approach in which the observed yields are parameterized in terms of the Wilson coefficients (WCs) of the EFT operators. A simultaneous fit of the 16 WCs to the data is performed and two standard deviation confidence intervals for the WCs are extracted; the standard model expectations for the WC values are within these intervals for all of the WCs probed.
}

\hypersetup{
pdfauthor={CMS Collaboration},
pdftitle={Search for new physics in top quark production with additional leptons in proton-proton collisions at sqrt(s) = 13 TeV using effective field theory},
pdfsubject={CMS},
pdfkeywords={CMS, physics, EFT}}

\maketitle

\section{Introduction}
\label{sec:intro}
There are numerous motivations to search for new particles or interactions
at the CERN LHC.  The strong evidence for dark matter~\cite{Feng:2010gw,Porter:2011nv} and dark energy~\cite{Peebles:2002gy} suggests the possibility that
the full list of nature's constituents has not yet been discovered.
Likewise, the observed asymmetry between matter and
antimatter in the universe implies an additional source of CP violation~\cite{Zyla:2020zbs}.
Many explanations for the hierarchy between the Planck mass and the electroweak 
scale set by the vacuum expectation value of the Higgs field also include new particles~\cite{Witten:1981nf,Alimena:2019zri,Matsedonskyi_2013}.
Nonetheless, there is no guarantee that new particles exist in the
mass range directly accessible at the LHC.  To extend the discovery
reach of the LHC, it is therefore important to consider not only direct
searches for new particles, but also indirect means of probing higher
energy scales.

One flexible framework for undertaking such indirect probes is that of
effective field theory (EFT)~\cite{Buchmuller:1985jz,Grzadkowski_2010,Falkowski:2019tft}. An EFT is a low-energy
approximation for a more fundamental theory involving interactions at a
mass scale $\Lambda$.
Treating the standard model (SM) as a low-energy approximation
of a higher-energy theory, one can add additional higher-order terms to 
the Lagrangian consistent with the symmetries and conservation laws that 
expand the possibilities compatible with new physics at higher mass scales.
The additional terms are constructed from 
operators---products of fields and their derivatives---that involve only SM fields.
The EFT Lagrangian can then be written as
\begin{linenomath}
\begin{equation} \label{effL}
\mathcal{L}_{\mathrm{EFT}} = \mathcal{L}_{\mathrm{SM}} + \sum\limits_{d,i} \dfrac{c^{d}_i}{\Lambda^{d-4}} \mathcal{O}^{d}_i ,
\end{equation}
\end{linenomath}
where $\mathcal{L}_{\mathrm{SM}}$ is the SM Lagrangian, $\mathcal{O}^{d}_i$ are the EFT
operators of dimension $d$, and $c^{d}_i$ are dimensionless parameters known as Wilson
coefficients (WCs) that characterize the strength of the interactions at dimension $d$.
The contribution of an operator of dimension $d$ to the Lagrangian is
suppressed by a factor of $1/\Lambda^{d-4}$ implying that the focus should
be placed on operators of the lowest possible dimension.  However,
all operators of odd dimension violate baryon or lepton number~\cite{Degrande:2012wf},
so dimension-five operators are neglected, and dimension-six operators provide
the leading contribution from new physics~\cite{Grzadkowski_2010}.
The first sub-leading contributions that conserve baryon and lepton number 
arise from dimension-eight operators and are not considered in this analysis.

While the impact of EFT operators can in principle be detected in a large
variety of experimental observables, it is particularly interesting to consider their
impact on the production of one or more top quarks in
association with additional leptons.  In the SM, the leading
contribution to this signature arises from the production of top quarks
in association with a \PW, \cPZ, or Higgs boson (\PH)~\cite{Zyla:2020zbs}.  It has long been
speculated that the large mass of the top quark~\cite{Abe:1995hr, D0:1995jca}, and hence its large
coupling to the Higgs boson, might be an indication of a special relationship
between the top quark and the physics of electroweak symmetry
breaking.  If so, the production of top quarks along with electroweak
or Higgs bosons may shed some light on possible new dynamics.  Only recently
have experimental measurements started to test directly the coupling
of the top quark to \PH~\cite{Sirunyan:2018hoz, Aaboud:2018urx} and \cPZ~\cite{Sirunyan:2017uzs, Sirunyan:2018zgs,
Aaboud:2019njj}
bosons.  The current and future LHC data therefore provide
intriguing opportunities to study these processes in more detail.
Furthermore, because of the new terms added within the EFT expansion, it is possible to consider the
production of top quarks with additional leptons directly through
four-fermion operators that do not contain \PH, \PW, or \cPZ bosons.  
Such contributions are also probed as part of this analysis.

Collisions producing one or more top quarks and additional leptons
generate a variety of signatures involving multiple leptons and jets,
including jets that are initiated from the hadronization of bottom quarks, referred to as  \PQb jets.  Top quarks decay with almost 100\%
branching fraction to a bottom quark and a \PW boson~\cite{Zyla:2020zbs}, which can
decay either leptonically, to a charged lepton and a neutrino, or
hadronically, to two jets.  If \PH, \PW, or \cPZ bosons are
produced in association with the top quarks, they may also decay in
various ways involving quarks (including  \PQb quarks, especially in the case of
the Higgs boson) or leptons.  Ultimately, the final-state signatures
are primarily determined by the decay modes of the bosons, either
hadronic or leptonic.  Final states in which multiple bosons
decay leptonically present a number of experimental advantages.
Multiple leptons provide an efficient trigger strategy, which remains
viable even at large instantaneous luminosities.  Furthermore, for final
states involving either a same-charge dilepton pair or more than two
leptons with additional jets (including  \PQb jets), the contributions from background processes are small compared to the size of the signals. These final states are the focus of this analysis and are denoted multilepton final states hereafter.

Focusing on multilepton final states leads to unique challenges that have not been encountered by previous
LHC analyses employing EFT methods to
search for new physics associated with top quark production~\cite{Sirunyan:2017uzs, Sirunyan:2018ucr, Sirunyan:2019wka,
CMS:2018jcg, CMS:2019too, Aad:2019pxo, Aaboud:2018nyl, Aaboud:2019njj,
Buckley:2015lku, Hartland:2019bjb, Brivio:2019ius}.
First, multilepton final states receive contributions
from multiple processes, and it is not possible to isolate
high-purity samples from each contribution.  For example, both \ttZ
and electroweak \tZq events contribute to the three-lepton final
state where two of the leptons form a same-flavor, oppositely charged pair
with an invariant mass near the \cPZ boson mass peak.  Likewise, same-charge
dilepton and trilepton final states outside the \cPZ peak originate
with comparable probability from SM \ttW and \ttH
production.  Since the multiple processes cannot be reliably
disentangled, this analysis cannot be constructed as a
reinterpretation of either a set of inclusive or differential cross
section measurements.  Second, there are numerous EFT
operators capable of impacting one or more of the processes contributing to multilepton final states;
a priori, there is no reason
to assume that new physics would manifest only through the
contribution of a single operator.
It is therefore important to analyze the effects of these operators simultaneously
across all components of the data set.

A new approach is implemented to address these challenges. Designed to target EFT effects directly, this approach does not aim to isolate specific physical processes and extract high-level observables; rather, it relies on detector-level observables, namely the number of events observed in a set of distinct categories defined by the multiplicities of final-state objects. For each category, a different admixture of physics processes will contribute to the observed event yield. Sensitivity to the EFT operators is obtained by parameterizing the predicted yields in terms of the WCs of all relevant operators simultaneously. To procure these predicted yields, we use simulated events with weights parameterized to represent the effects of the EFT operators. These weighted, simulated events are then analyzed to obtain the necessary predictions of the observed event yields, as functions of the EFT parameters.
Parameterizing the event weights in terms of the WCs represents the key enabling concept of this approach, as it allows all relevant interference effects---both interference between new physics and the SM and interference among new physics operators---to be incorporated into the prediction. The effects of multiple EFT operators on multiple physical processes contributing to a single final-state signature are therefore accounted for in a straightforward and rigorous manner. EFT operators can also impact the kinematical properties of the events, so this approach allows the full effect on the detector acceptance and efficiency to be appropriately described. Correlations among statistical and systematic uncertainties can also be accounted for, and, where possible, fully leveraged. For example, this approach should provide enhanced sensitivity when EFT operators impact the contribution of multiple relevant physics processes, since the observables used are sensitive to the sum of the effects.
The main drawback of this approach is that, because it relies on detector-level observation and fully simulated events, theoretical updates cannot be incorporated without repeating the analysis. 
This is the first time such an approach has been applied to LHC data; ultimately, the technique can be applied to differential kinematical distributions, but for this initial analysis, we take a more inclusive approach.

The detailed strategy employed in this analysis is as follows.
Multilepton events are divided into categories based on the number and
the sign of the charge sum of the reconstructed leptons; the lepton categories are
then subdivided according to the number of  \PQb jets. Within each
lepton and  \PQb jet category, the event yields are characterized as a
function of the number of jets.  For oppositely charged,
same-flavor lepton pairs in three-lepton events, the data are divided based on whether
the invariant mass of the lepton pair falls in a window around the \cPZ boson mass ($m_{\cPZ}$).
This strategy results in 35 nonoverlapping categories.  These
event yields define the observables for the analysis and are compared
against predictions that incorporate the effects of EFT operators.
Contributions involving primarily prompt leptons---including
signal processes---are modeled using simulated events.  Where
relevant, the predicted yields for processes sensitive to EFT operators
are parameterized in terms of the WCs for those
operators.  Predictions for backgrounds involving primarily
nonprompt leptons (\eg, leptons from bottom or charmed hadron decays
or misidentified leptons) are based on extrapolations from control regions
in data.  The WCs are varied to
determine the best fit of the predictions to data, as well as to
establish the range over which the predicted yields are consistent with the observation.

The sections of the paper are organized in the following order. A brief overview of the CMS detector and triggering system is outlined in Section~\ref{sec:detector}. Section~\ref{sec:samples} describes the simulation of signal and background processes, including a discussion of the parameterization of the predicted yields in terms of the WCs. The event reconstruction and event selection are covered in Sections~\ref{sec:reco} and~\ref{sec:selection}, respectively, while Section~\ref{sec:backgrounds} discusses the background estimation. In Section~\ref{sec:fitting}, the signal extraction is explained. Sources of systematic uncertainties affecting this analysis are described in Section~\ref{sec:systematics}. Section~\ref{sec:results} presents the results, and Section~\ref{sec:summary} provides a summary of the analysis.

\section{The CMS detector}
\label{sec:detector}

The central feature of the CMS apparatus is a superconducting solenoid of 6\unit{m} internal diameter, providing a magnetic field of 3.8\unit{T}.  Within the solenoid volume are a silicon pixel and strip tracker, a lead tungstate crystal electromagnetic calorimeter (ECAL), and a brass and scintillator hadron calorimeter (HCAL), each composed of a barrel and two endcap sections.  Forward calorimeters extend the pseudorapidity coverage provided by the barrel and endcap detectors.  Muons are detected in gas-ionization chambers embedded in the steel flux-return yoke outside the solenoid.  A more detailed description of the CMS detector and its performance, together with a definition of the coordinate system and the kinematic variables used in the analysis, can be found in Ref.~\cite{Chatrchyan:2008zzk}.

Events of interest are selected using a two-tiered trigger system~\cite{Khachatryan:2016bia}.  The first level, composed of custom hardware processors, uses information from the calorimeters and muon detectors to select events at a rate of around 100\unit{kHz} within a fixed time interval of about 4\mus.  The second level, known as the high-level trigger, consists of a farm of processors running a version of the full event reconstruction software optimized for fast processing, and reduces the event rate to around 1\unit{kHz} before data storage.

\section{Data samples and simulation}
\label{sec:samples}

The data used in this analysis comprise proton-proton ($\Pp\Pp$) collisions at $\sqrt{s} = 13\TeV$ collected with the CMS detector in 2017, corresponding to a total integrated luminosity of \lumi~\cite{LUM-17-004}.
The events have been recorded using a combination of single-, double-, and triple-lepton triggers.

Simulations are used to estimate the event yields of the signal processes and some sources of background.
The signal samples incorporate EFT effects and are generated at leading order (LO), while all background samples are generated at next-to-leading order (NLO) and do not include  EFT effects.
The simulated samples used to estimate the backgrounds include $ \PQt\cPaqt\PGg$, diboson, and triboson production.
Used for validation purposes, additional samples are also generated to simulate SM background processes that are estimated from data;
these include \ZJets, \WJets, \ttbarJets, and single top quark processes ($s$ channel, $t$ channel, and \tW).
The background samples are generated using matrix elements (MEs) implemented either in the \MGvATNLO~\cite{MadGraph5_aMCatNLO,Frederix:2012ps,Alwall:2007fs} (version 2.4.2) or the \POWHEG v2~\cite{POWHEG1,POWHEG2,POWHEG3,POWHEG_ST_tch_4f,POWHEG_ST_tW,POWHEG_VV,POWHEG_TTbar} programs.
The simulation of the signal processes is described in Section~\ref{sec:signal_sample_generation}

Parton showering and hadronization for all of the samples is done by \PYTHIA~\cite{pythia8} (version 8.226 was used for the signal samples), and the Lund fragmentation model is employed~\cite{PYTHIA_MonashTune}.
The parameters for the underlying event description correspond to the CP5 tune~\cite{Sirunyan:2019dfx} and the proton structure is described by the NNPDF3.1~\cite{NNPDF3} set of parton distribution functions (PDFs).
Minimum bias $\Pp\Pp$ interactions occurring in the same or nearby bunch crossings (pileup) are generated with \PYTHIA and overlaid on all simulated events, according to the luminosity profile of the analyzed data.  
Finally, all generated events are passed through a detailed simulation of the CMS apparatus, based on \GEANTfour~\cite{geant4}, and are reconstructed using the same event reconstruction software as used for data.

\subsection{Simulation of the signal processes}
\label{sec:signal_sample_generation}

\begin{figure}[!th]
\centering
\includegraphics[height=0.2\textwidth]{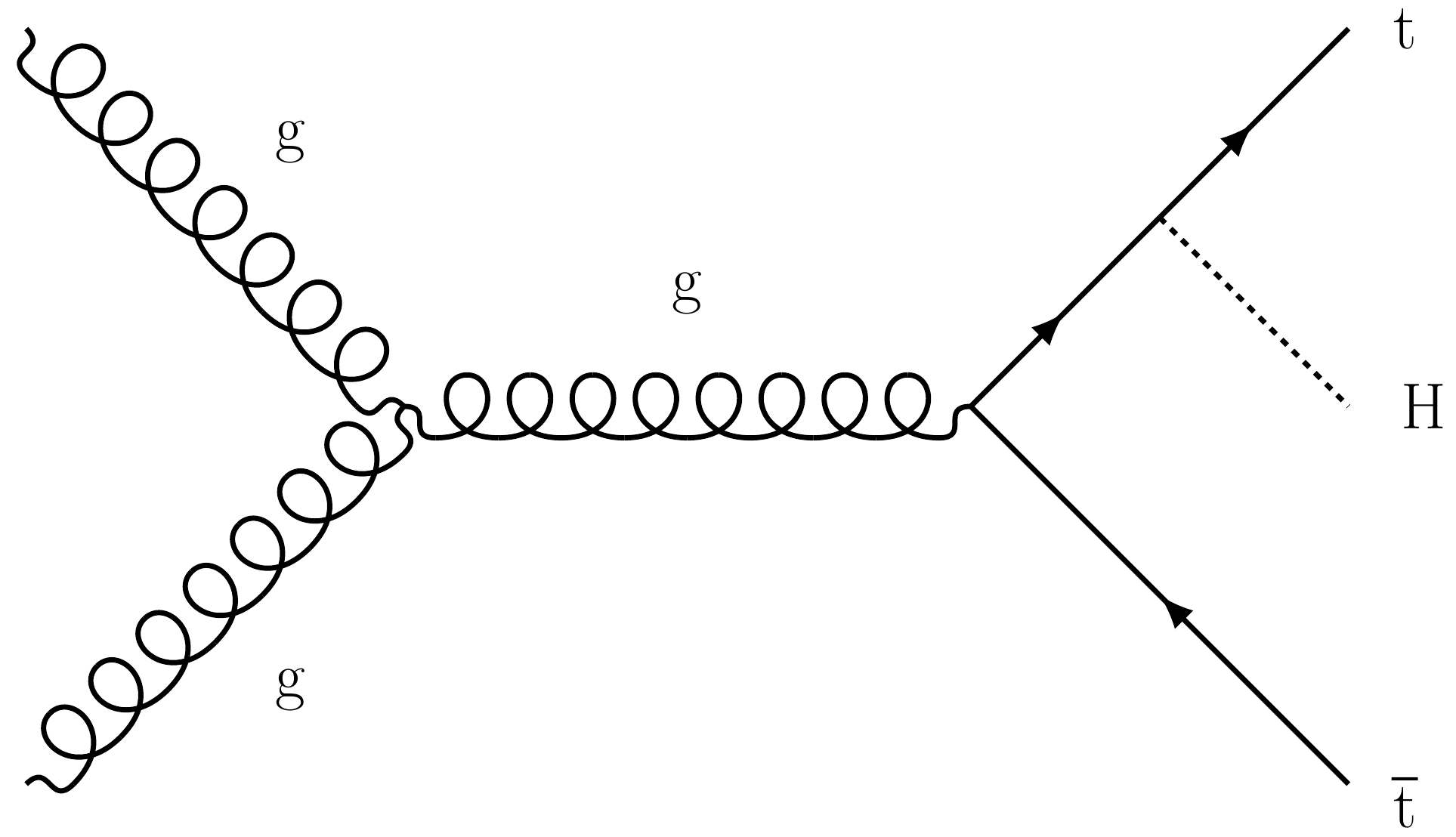} \hskip 5ex
\includegraphics[height=0.2\textwidth]{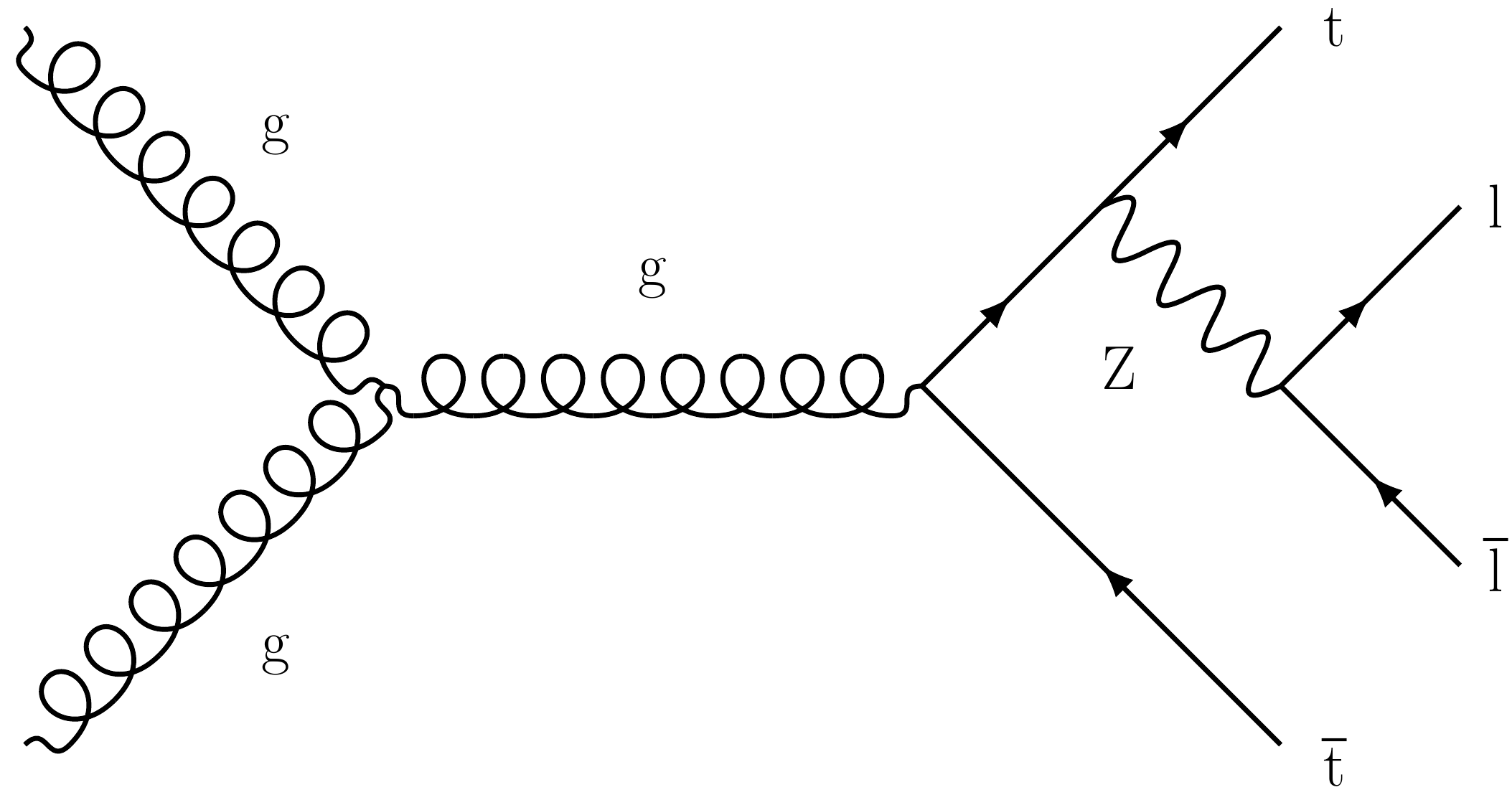} \\ \vskip 5ex
\includegraphics[width=0.35\textwidth]{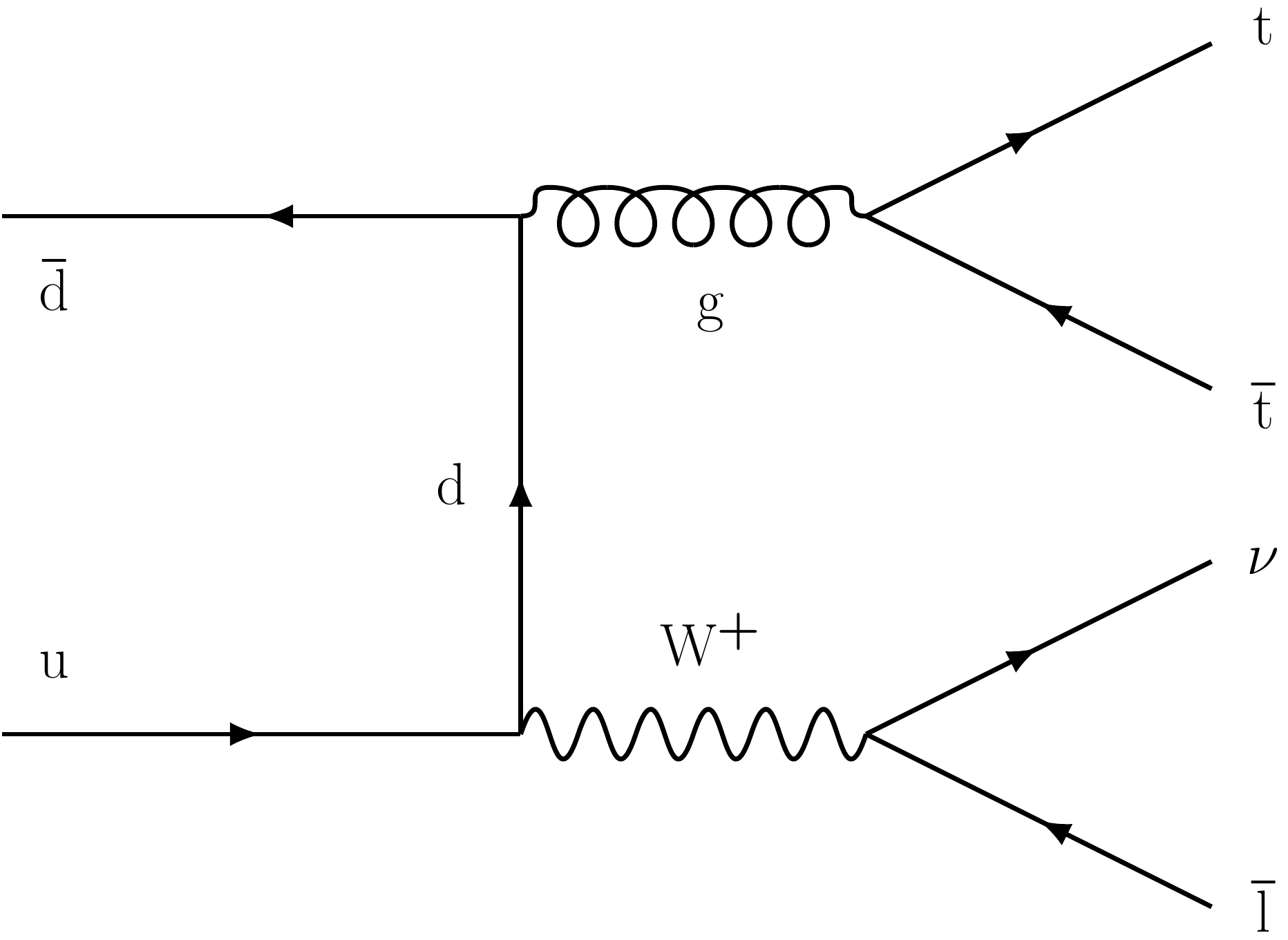} \\ \vskip 5ex
\includegraphics[height=0.15\textwidth]{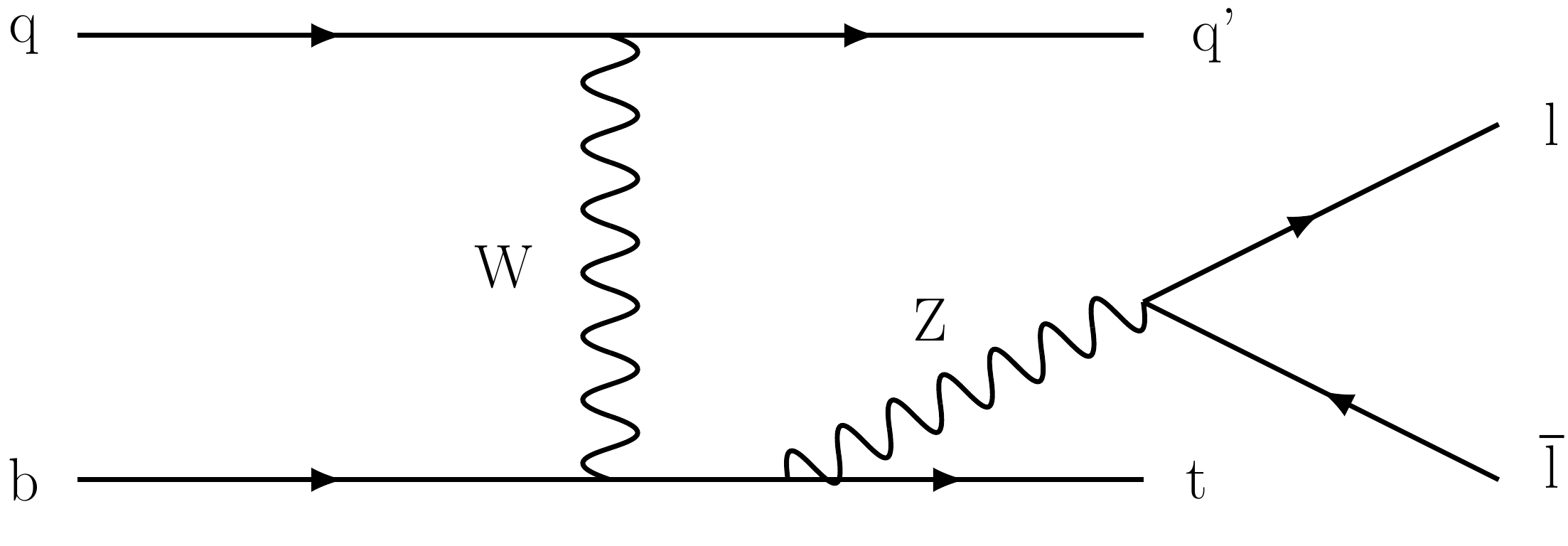} \hskip 5ex 
\includegraphics[height=0.15\textwidth]{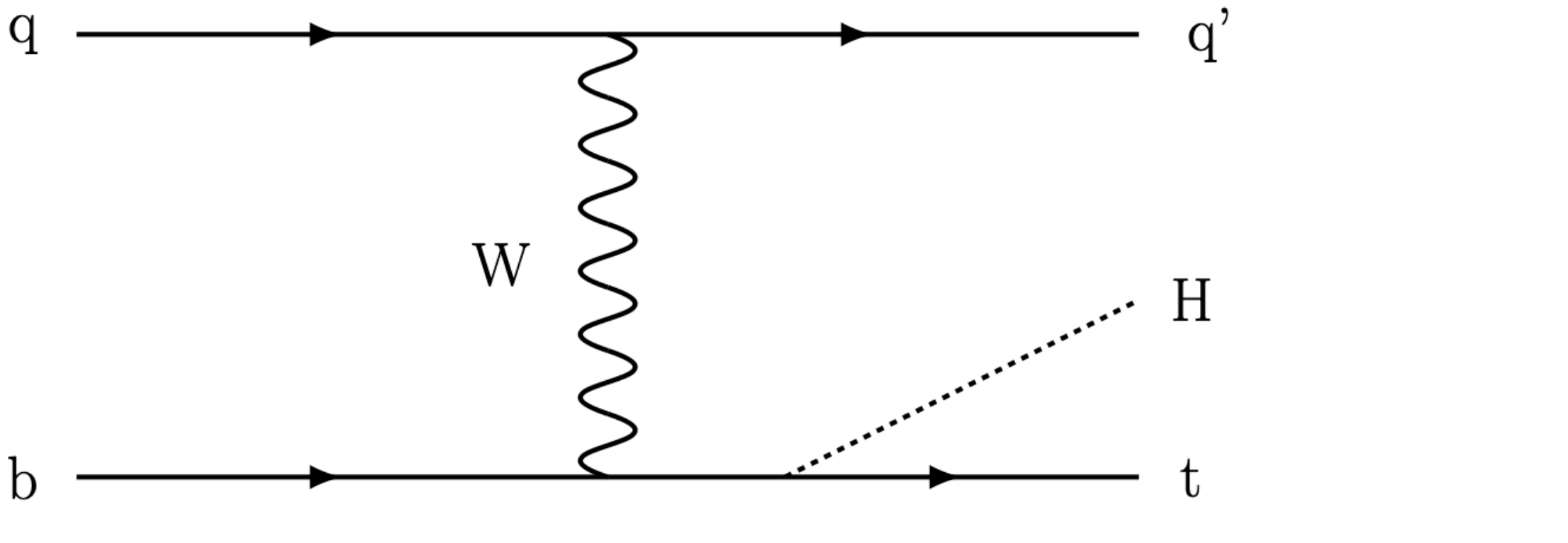} \vskip 5ex
\caption{Example diagrams for the five signal processes considered in this analysis: \ttH, \ttll, \ttlnu, \tllq, and \tHq.}
\label{fig:signal_proc_feynman_dias}
\end{figure}

The signal events are generated at LO with \MGvATNLO (version 2.6.0).
The signal processes include those in which one or more top quarks are produced along with multiple charged leptons: \ttll, \tllq, and \ttlnu (where l indicates a charged lepton and $\nu$ indicates a neutrino). We also include \ttH and \tHq, as these processes can produce signal events when the Higgs boson decays into one or more leptons. The decays of the Higgs bosons are handled by \PYTHIA since it would be computationally expensive to produce \MGvATNLO samples for each decay mode and difficult to separate them from the other signal processes.
An example diagram for each signal process is shown in Fig.~\ref{fig:signal_proc_feynman_dias}.
We note that the signal processes include contributions from lepton pairs produced from on-shell \PW and \cPZ bosons, as well as those from nonresonant processes; this is important so that effects from EFT four-fermion operators can also be included in these samples.
Furthermore, we note that the \ttll sample includes the production of top quark pairs in association with virtual photons.

\begin{figure}[!t]
\centering
\includegraphics[height=0.2\textwidth]{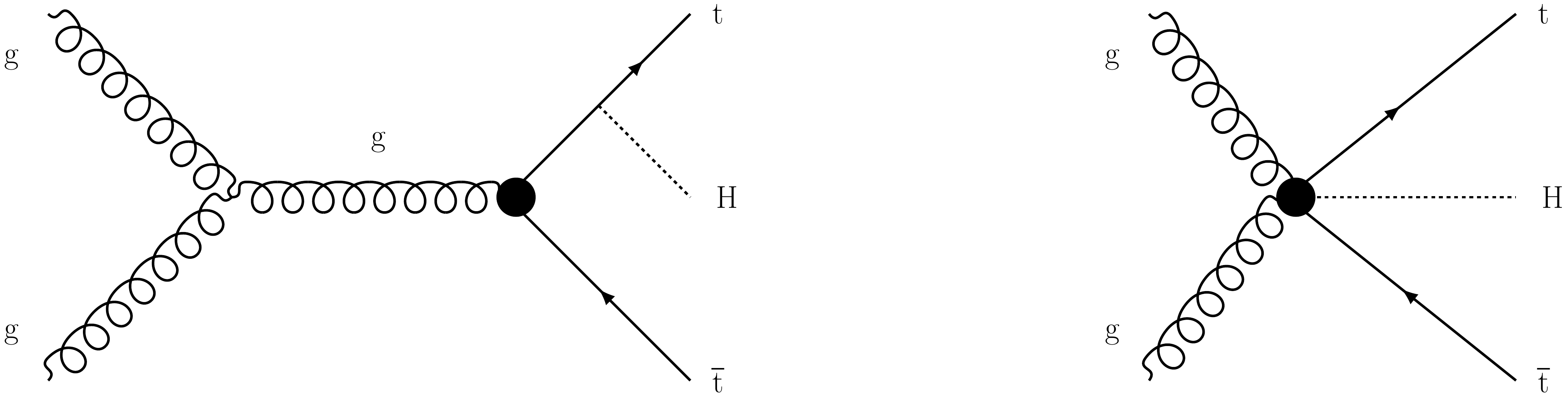} \hskip 5ex
\caption{Example diagrams showing two of the vertices associated with the \qq{}{ \PQu G}{} operator. This operator, whose definition can be found in Table~\ref{tab:eftOperators}, gives rise to vertices involving top quarks, gluons, and the Higgs boson; as illustrated here, these interactions can contribute to the \ttH process.}
\label{fig:ctG_feynman_dias}
\end{figure}

Sixteen operators are considered in this analysis. Chosen because they are expected to have a relatively large impact on the signal processes but not on the \ttbar background process, the operators can be classified into two categories: operators that involve two quarks and one or more bosons, and operators that involve two quarks and two leptons.  Table~\ref{tab:eftOperators} lists the 16 operators and corresponding WCs; all couplings are assumed to involve only quarks of the third generation.  The operators that require a Hermitian conjugate term in the Lagrangian (marked with a double dagger in Table~\ref{tab:eftOperators}) can have complex WCs; however, the imaginary coefficients lead to CP violation, and, as outlined in Ref.~\cite{AguilarSaavedra:2018nen}, are generally already constrained.  Consequently, only the real components are considered in this analysis.
More details about the operators listed in Table~\ref{tab:eftOperators} can be found in Ref.~\cite{AguilarSaavedra:2018nen}.
For the purpose of illustrating which processes are most strongly affected by the operators considered, Table~\ref{tab:eftOperators} lists the leading signal processes affected by each operator. 
To determine whether a process is to be listed in this column, we check whether the cross section of the process is scaled by more than five times the SM cross section when the WC associated with the operator is set to $16\pi^2$ or $-16\pi^2$~\cite{Degrande:2012wf}.
For the operators that are associated with multiple WCs, if any of the WCs causes the process to be scaled by more than five times the SM cross section, the process is considered to be affected by that operator.
The choice to use a factor of five as the criterion for this determination is arbitrary; if it were changed, the list of processes listed in the column  would potentially change as well.
However, changing these criteria would have no influence on the analysis, as the effects of all operators on all process are fully considered.
The diagrams in Fig.~\ref{fig:ctG_feynman_dias} illustrate two examples of interactions that arise from one of the operators we consider; these interactions can affect the signal process \ttH.

\begin{table}[hbt!]
\topcaption{List of operators that have effects on \ttH, \ttll, \ttlnu, \tllq, and \tHq processes at order $1/\Lambda^2$ that are considered in this analysis. The couplings are assumed to involve only third-generation quarks. 
The quantity $T^A=\frac{1}{2}\lambda^A$ denotes the eight Gell-Mann matrices, and $\tau^I$ are the Pauli matrices.
The field $\varphi$ is the Higgs boson doublet, and $\tilde{\varphi}=\varepsilon\varphi^*$, where $\varepsilon \equiv i\tau^2$.
The $\ell$ and \cPq represent the left-handed lepton and quark doublets, respectively, while \Pe represents the right-handed lepton, and  \PQu and  \PQd represent the right-handed quark singlets.  
Furthermore,
    $(\varphi^{\dagger}i\protect\overleftrightarrow{D}_\mu\varphi)    \equiv \varphi^\dagger(iD_\mu\varphi)-(iD_\mu\varphi^\dagger)\varphi $ and
    $(\varphi^{\dagger}i\protect\overleftrightarrow{D}^I_\mu\varphi)  \equiv \varphi^{\dagger}\tau^I(iD_\mu\varphi)-(iD_\mu\varphi^\dagger)\tau^I\varphi $.
The \PW boson field strength is $\PW_{\mu\nu}^I=\partial_\mu \PW_\nu^I-\partial_\nu \PW^I_\mu+g\varepsilon_{IJK}\PW^J_\mu \PW^K_\nu$, and $G_{\mu\nu}^A=\partial_\mu G_\nu^A-\partial_\nu G^A_\mu+g_sf^{ABC}G^B_\mu G^C_\nu$ is the gluon field strength.  
The abbreviations \sW and \cW denote the sine and cosine of the weak mixing angle (in the unitary gauge), respectively.
The leading processes affected by the operators are also listed (the details of the criteria used for this determination are described in the text).
}
\begin{center}
\begin{tabular}{p{1.5cm}p{3.7cm}p{3.5cm}p{4.1cm}}
\hline
\multicolumn{4}{c}{Operators involving two quarks and one or more bosons \rule[2.5ex]{0pt}{0pt} \rule[-1.0ex]{0pt}{0pt}} \\ 
Operator                     & Definition                                                             & WC                       & Lead processes affected\rule[2.5ex]{0pt}{0pt} \rule[-1.0ex]{0pt}{0pt} \\ 
\hline
$\hc{\qq{}{ \PQu\varphi}{ij}}$ \rule{0pt}{2.6ex}       & $\PAQq_i  \PQu_j\tilde\varphi\: (\varphi^{\dagger}\varphi)$             & $ \ctp + i \ctpI     $   & \ttH, \tHq                       \\ 
$\qq{1}{\varphi \cPq}{ij}$           & $\FDF (\PAQq_i\gamma^\mu \cPq_j)$                                      & $ \cpQM + \cpQa      $   & \ttH, \ttlnu, \ttll, \tHq, \tllq \\ 
$\qq{3}{\varphi \cPq}{ij}$           & $\FDFI (\PAQq_i\gamma^\mu\tau^I \cPq_j)$                               & $ \cpQa              $   & \ttH, \ttlnu, \ttll, \tHq, \tllq \\ 
$\qq{}{\varphi  \PQu}{ij}$            & $\FDF (\PAQu_i\gamma^\mu  \PQu_j)$                                      & $ \cpt               $   & \ttH, \ttlnu, \ttll, \tllq       \\ 
$\hc{\qq{}{\varphi  \PQu \PQd}{ij}}$   & $(\tilde\varphi^\dagger iD_\mu\varphi)(\PAQu_i\gamma^\mu  \PQd_j)$      & $ \cptb + i \cptbI   $   & \ttH, \tllq, \tHq                \\ 
$\hc{\qq{}{ \PQu\PW}{ij}}$            & $(\PAQq_i\sigma^{\mu\nu}\tau^I \PQu_j)\:\tilde{\varphi}\PW_{\mu\nu}^I$  & $ \ctW + i \ctWI     $   & \ttH, \ttlnu, \ttll, \tHq, \tllq \\ 
$\hc{\qq{}{ \PQd\PW}{ij}}$            & $(\PAQq_i\sigma^{\mu\nu}\tau^I \PQd_j)\:{\varphi} \PW_{\mu\nu}^I$       & $ \cbW + i \cbWI     $   & \ttH, \ttll, \tHq, \tllq         \\ 
$\hc{\qq{}{ \PQu\PB}{ij}}$            & $(\PAQq_i\sigma^{\mu\nu} \PQu_j)\:\tilde{\varphi}\PB_{\mu\nu}$     & $ (\cW \ctW-\ctZ)/\sW + i (\cW \ctWI - \ctZI)/\sW $ & \ttH, \ttlnu, \ttll, \tHq, \tllq \\
$\hc{\qq{}{ \PQu G}{ij}}$ \rule[-2.5ex]{0pt}{0pt}            & $(\PAQq_i\sigma^{\mu\nu}T^A \PQu_j)\:\tilde{\varphi}G_{\mu\nu}^A$       & $ \mathrm{g}_\mathrm{s}(\ctG + i \ctGI)     $   & \ttH, \ttlnu, \ttll, \tHq, \tllq \\ \hline
\multicolumn{4}{c}{Operators involving two quarks and two leptons \rule[2.5ex]{0pt}{0pt} \rule[-1.0ex]{0pt}{0pt}} \\ 
Operator                                    & Definition                                                                                  & WC                    & Lead processes affected\rule[2.5ex]{0pt}{0pt} \rule[-1.0ex]{0pt}{0pt}\\ 
\hline
$\qq{1}{\ell\cPq}{ijkl}$ \rule{0pt}{2.6ex}                &$(\PAlepton_i\gamma^\mu \ell_j)(\PAQq_k\gamma^\mu \cPq_\ell)$                        & $ \cQlM + \cQla     $ &  \ttlnu, \ttll, \tllq \\ 
$\qq{3}{\ell\cPq}{ijkl}$                &$(\PAlepton_i\gamma^\mu \tau^I \ell_j)(\PAQq_k\gamma^\mu \tau^I \cPq_\ell)$          & $ \cQla             $ &  \ttlnu, \ttll, \tllq \\ 
$\qq{}{\ell \PQu}{ijkl}$                 &$(\PAlepton_i\gamma^\mu \ell_j)(\PAQu_k\gamma^\mu  \PQu_\ell)$                        & $ \ctl              $ &  \ttll                \\ 
$\qq{}{\Pe\PAQq}{ijkl}$                     &$(\overline \Pe_i\gamma^\mu \Pe_j)(\PAQq_k\gamma^\mu \cPq_\ell)$                              & $ \cQe              $ &  \ttll, \tllq         \\ 
$\qq{}{\Pe \PQu}{ijkl}$                      &$(\overline \Pe_i\gamma^\mu \Pe_j)(\PAQu_k\gamma^\mu  \PQu_\ell)$                              & $ \cte              $ &  \ttll                \\ 
$\hc{\qq{1}{\ell\Pe\cPq \PQu}{ijkl}}$    &$(\PAlepton_i \Pe_j)\;\varepsilon\;(\PAQq_k  \PQu_\ell)$                                  & $ \ctlS + i \ctlSI  $ &  \ttll, \tllq         \\ 
$\hc{\qq{3}{\ell\Pe\cPq \PQu}{ijkl}}$ \rule[-1.5ex]{0pt}{0pt}   &$(\PAlepton_i \sigma^{\mu\nu} \Pe_j)\;\varepsilon\;(\PAQq_k \sigma_{\mu\nu}  \PQu_\ell)$  & $ \ctlT + i \ctlTI  $ &  \ttlnu, \ttll, \tllq \\ 
\hline
\end{tabular}
\label{tab:eftOperators}
\end{center}
\end{table}

The signal events are generated using an approach similar to the one described in Ref.~\cite{AguilarSaavedra:2018nen}. Like the framework presented there, the model used in this analysis adopts the Warsaw basis of gauge-invariant dimension-six operators~\cite{Grzadkowski_2010}, focuses on operators that give rise to interactions involving at least one top quark, and only allows for tree-level generation. 
To allow \MGvATNLO to properly handle the emission of gluons from \qq{}{ \PQu G}{} vertices, an extra factor of the strong coupling is applied to the coefficients arising from the \qq{}{ \PQu G}{} operator, as indicated in Table~\ref{tab:eftOperators}.
Since only tree-level generation is possible with this model, the \ttH, \ttlnu, and \ttll signal samples are generated with an extra parton in the final state, improving accuracy and allowing some operators to contribute more significantly to processes upon which they would otherwise have a weaker effect.

For the samples generated with an additional parton, a matching procedure is applied to account for the overlap in phase space between the ME and parton shower (PS) contributions; in this analysis, we use the MLM scheme~\cite{Alwall:2007fs}. Since EFT effects are included in the ME contribution but not in the PS contribution, the validity of the matching procedure must be examined for operators that give rise to vertices involving a gluon. The only such operator considered by this analysis is \qq{}{ \PQu G}{}, and contributions from this operator to the soft and collinear regime are suppressed; therefore, the phase space overlap with the SM contribution from the PS is small, and the effects of this potential issue are mitigated~\cite{Goldouzian:2020ekx}.
The \tllq and \tHq signal samples are not generated with an extra final state parton since, when using the MLM scheme with LO \MGvATNLO, jet matching cannot be correctly performed between the ME and the PS for single top quark processes in the $t$ channel.

\subsection{Parameterization of the yields in terms of the WCs}
\label{sec:parametrization}
In order to discern the effects of new physics on the observed yields, the predicted yields must be parameterized in terms of the WCs. First, the ME can be written as the sum of SM and new physics components:
\begin{linenomath}
\begin{equation}
\mathcal{M} = \mathcal{M}_{\mathrm{SM}} + \sum_{i} \frac{c_{i}}{\Lambda^2}\mathcal{M}_{i},
\label{eq:EFTMatrixElement}
\end{equation}
\end{linenomath}
where $\mathcal{M}_{\mathrm{SM}}$ is the SM ME, $\mathcal{M}_{i}$ are the MEs corresponding to the new physics components, and $c_{i}$ are the WCs. Any cross section---inclusive or differential---is proportional to the square of the ME.
The SM contribution to the cross section is necessarily independent of the WC values, while the new EFT operators have contributions that depend linearly on the coefficients arising from the interference with the SM and contributions quadratic in the coefficients from pure EFT terms. The latter can originate from the effects of an individual operator or interference between the EFT operators.
Since this analysis considers 16 WCs, the expected cross section could therefore be parameterized as a 16-dimensional (16D) quadratic function of the WCs.

In principle, the 16D quadratic terms could be fully determined by evaluating the cross section at multiple points in WC space and solving for the coefficients; it would be impractical, however, to produce the large number of simulated samples required. Instead, we use the \MGvATNLO event generator's ability to assign weights to the generated events in order to effectively account for the variation of the differential cross section in an infinitesimal part of phase space occupied by an individual event. 
Each event weight, just like the inclusive or differential cross section, can be parameterized by a polynomial of second order in the WCs. In order to determine the coefficients of each event's 16D quadratic weight function, the weight is calculated at multiple randomly selected points in the 16D WC phase space. Once enough weights have been calculated to constrain the 16D quadratic function, we can solve for the coefficients and therefore obtain the parameterization for the weight function of each event in terms of the WCs. The weight functions $w_i$ for a given event event $i$ can then be written as follows:
\begin{linenomath}
\begin{equation}
w_i\left(\frac{\vec{c}}{\Lambda^2}\right) = s_{0i} + \sum_js_{1ij}\frac{c_j}{\Lambda^2} + \sum_js_{2ij}\frac{c_j^2}{\Lambda^4} + \sum_{j,k}s_{3ijk} \frac{c_j}{\Lambda^2} \frac{c_k}{\Lambda^2},
\label{eq:quad_wgt_eqn}
\end{equation}
\end{linenomath}
where $\vec{c}$ represents the set of WCs, the sum over $j$ and $k$ corresponds to the sum over the WCs, and the values $s_{0}$, $s_{1}$, $s_{2}$, and $s_{3}$ represent the coefficients in the quadratic parameterization of the weight from the SM, interference between EFT and SM, pure EFT, and interference between two EFT contributions, respectively.

The weighted events can then be used to calculate predicted yields for an arbitrary set of WC values;
the yield for a given event selection category (as discussed in Section~\ref{sec:selection}) is found by summing the weight functions for events that meet the selection requirements of the given category. Thus, summing Eq.~(\ref{eq:quad_wgt_eqn}) over $i$, we find the predicted yield $N$:
\begin{linenomath}
\begin{equation}
\begin{aligned}
N\left(\dfrac{\vec{c}}{\Lambda^2}\right) &= \sum_i w_i\left(\frac{\vec{c}}{\Lambda^2}\right) \\
           &= \sum_i \left(s_{0i} + \sum_js_{1ij} \frac{c_j}{\Lambda^2} + \sum_js_{2ij}\frac{c_j^2}{\Lambda^4} + \sum_{j,k}s_{3ijk} \frac{c_j}{\Lambda^2} \frac{c_k}{\Lambda^2} \right)\\
           &= \left(\sum_i s_{0i}\right) + \sum_j \left(\sum_i s_{1ij} \right)\frac{c_j}{\Lambda^2} + \sum_j \left(\sum_i s_{2ij}\right)\frac{c_j^2}{\Lambda^4} + \sum_{j,k} \left(\sum_i s_{3ijk}\right) \frac{c_j}{\Lambda^2}\frac{c_k}{\Lambda^2} \\
           &= S_0 + \sum_j S_{1j} \frac{c_j}{\Lambda^2} + \sum_j S_{2j} \frac{c_j^2}{\Lambda^4} + \sum_{j,k} S_{3jk} \frac{c_j}{\Lambda^2} \frac{c_k}{\Lambda^2}.
\end{aligned}
\end{equation}
\end{linenomath}
The predicted yield has therefore been expressed as a quadratic function of the WCs, where the quadratic coefficients of the yield parameterization were found by summing the quadratic coefficients of the weights, \eg, $S_{1j} = \sum_i s_{1ij}$. Since the parameterized yields should be consistent with the SM when all WCs are set to zero, we normalize the yields to the NLO predictions~\cite{deFlorian:2016spz}.

\section{Event reconstruction}
\label{sec:reco}

The CMS global (also called particle-flow (PF)~\cite{CMS-PRF-14-001}) event reconstruction aims to reconstruct and identify each individual particle in an event, with an optimized combination of all subdetector information.  In this process, the identification (ID) of the particle type (photon, electron, muon, charged or neutral hadron) plays an important role in the determination of the particle direction and energy.  Photons are identified as ECAL energy clusters not linked to the extrapolation of any charged particle trajectory to the ECAL.  Electrons are identified as a primary charged particle track and potentially many ECAL energy clusters corresponding to this track extrapolation to the ECAL and to possible bremsstrahlung photons emitted along the way through the tracker material.  Muons are identified as tracks in the central tracker consistent with either a track or several hits in the muon system, and associated with calorimeter deposits compatible with the muon hypothesis.  Charged hadrons are identified as charged-particle tracks neither identified as electrons, nor as muons.  Finally, neutral hadrons are identified as HCAL energy clusters not linked to any charged hadron trajectory, or as a combined ECAL and HCAL energy excess with respect to the expected charged hadron energy deposit.

The energy of photons is obtained from the ECAL measurement.  The energy of electrons is determined from a combination of the charged track momentum at the main interaction vertex, the corresponding ECAL cluster energy, and the energy sum of all bremsstrahlung photons attached to the track.  The energy of muons is obtained from the corresponding track momentum.  The energy of charged hadrons is determined from a combination of the track momentum and the corresponding ECAL and HCAL energies, corrected for the response function of the calorimeters to hadronic showers.  Finally, the energy of neutral hadrons is obtained from the corresponding corrected ECAL and HCAL energies.

The missing transverse momentum vector (\ptvecmiss) is computed as the negative vector sum of the transverse momenta of all the PF candidates in an event, and its magnitude is denoted as \ptmiss~\cite{Sirunyan:2019kia}.  The vector \ptvecmiss is modified to account for corrections to the energy scale of the reconstructed jets in the event. 

The candidate vertex with the largest value of the sum of squared physics-object transverse momentum (\pt) is taken to be the primary $\Pp\Pp$ interaction vertex.  The physics objects are the jets, clustered using the anti-\kt jet finding algorithm~\cite{Cacciari:2008gp,Cacciari:2011ma} with the tracks assigned to the vertex as inputs, and the associated \ptvecmiss.  More details are given in Section~9.4.1 of Ref.~\cite{CMS-TDR-15-02}.  Lepton candidates, which are subsequently reconstructed, are required to be compatible with originating from the selected primary vertex.

Electrons are reconstructed by matching tracks in the silicon tracker to the energy deposit in the ECAL, without any significant energy deposit in the HCAL~\cite{Khachatryan:2015hwa}.  Genuine electrons are distinguished from hadrons mimicking an electron signature by a multivariate algorithm using the quality of the electron track, the shape of the electron cluster, and the matching between the track momentum and direction with the sum and position of energy deposits in the ECAL.  Furthermore, to reject electrons produced in photon conversions, candidates with missing hits in the innermost tracking layers or matched to a conversion secondary vertex are discarded.

Muon candidates are reconstructed by combining information from the silicon tracker and the outer muon spectrometer of CMS in a global fit~\cite{Sirunyan:2018fpa}.  The quality of the geometrical matching between the individual measurements in the tracker and the muon system is used to improve the ID of genuine prompt muons by substantially reducing misidentification due to hadrons punching through the calorimeters or from muons produced through in-flight decays of kaons and pions.

The electron and muon selection criteria described above define the ``loose'' lepton selection.  Additional selection criteria are applied to discriminate leptons produced in the decays of \PW and \cPZ bosons and \Pgt leptons from leptons produced in the decays of  \PQb or light hadrons, or from misidentified jets.  We will refer to the former as ``prompt'' leptons and to the latter as ``nonprompt'' leptons.  Isolation criteria are also applied to all leptons.

A multivariate analysis (MVA) approach based on boosted decision trees (BDTs), referred to as the prompt lepton MVA, is used for this selection.  Each BDT takes as inputs the lepton kinematic, ID, and impact parameter information, as well as kinematic and  \PQb tagging information about the nearest jet to the lepton.  Two versions of the lepton MVA are trained, one for electrons and one for muons, which differ based on the inclusion of one additional observable for each of the two versions.  For electrons, the extra input is the multivariate discriminant developed via dedicated analysis for electron ID~\cite{Khachatryan:2015hwa}.  For muons, the extra input is a similar discriminant that classifies the compatibility of track segments in the muon system with the pattern expected from a muon ionization~\cite{Chatrchyan:2012xi}.
The BDT inputs have been checked in control regions in data to ensure that they are well modeled by the simulation.
A more detailed description of the lepton MVA can be found in~\cite{HIG-17-018}.

In the signal region, lepton candidates are required to exceed a given discriminant threshold, referred to as the ``tight'' lepton selection.  A looser selection, referred to as the ``relaxed'' selection, is defined by relaxing the above lepton MVA discriminant requirement for the purpose of estimating background processes, as discussed in Section~\ref{sec:backgrounds}. The efficiency of the triggers based on the presence of one, two, or three electrons or muons is measured in data in $\cPZ/\Pggx \to \Pe\Pe$ and $\cPZ/\Pggx \to \Pgm\Pgm$ events, respectively. These events are also used to measure the efficiency for electrons or muons to pass the lepton reconstruction, ID, and isolation criteria~\cite{Khachatryan:2015hwa,Sirunyan:2018fpa}. 

For each event, hadronic jets are clustered from PF candidates using the anti-\kt algorithm with a distance parameter of 0.4.  Jet momentum is determined as the vectorial sum of all particle momenta in the jet, and is found from simulation to be, on average, within 5 to 10\% of the true momentum over the whole $\pt$ spectrum and detector acceptance.  Pileup can contribute additional tracks and calorimetric energy depositions to the jet momentum.  To mitigate this effect, charged particles originating from pileup vertices are discarded and an offset correction is applied to account for the remaining contributions.  Jet energy corrections are derived from simulation to bring measured response of jets to that of particle level jets on an average.  In situ measurements of the momentum balance in dijet, $\text{photon}{+}\text{jet}$, $\ZJets$, and multijet events are used to account for any residual differences in jet energy scale in data and simulation~\cite{Khachatryan:2016kdb}.  The jet energy resolution amounts typically to 15--20\% at 30\GeV, 10\% at 100\GeV, and 5\% at 1\TeV~\cite{Khachatryan:2016kdb}.  Jets are rejected if the distance in $\eta$--$\phi$ space (where the $\phi$ is azimuthal angle in radians) between the jet and the closest lepton is less than 0.4.

Jets originating from the hadronization of  \PQb quarks are identified by two ``combined secondary vertex'' algorithms~\cite{Chatrchyan:2012jua,BTV-16-002}, namely CSVv2 and DeepCSV, which exploit observables related to the long lifetime of  \PQb  hadrons and to the higher particle multiplicity and mass of  \PQb jets compared to light-quark and gluon jets.  DeepCSV is used for  \PQb jet selection, while CSVv2 is used to aid lepton ID as an input to the prompt lepton MVA.  The analysis uses two levels of selection based on DeepCSV, with a loose and a medium working point. The medium (loose) working point has approximately 70 (85)\% efficiency for tagging jets from  \PQb quarks, with a misidentification probability of $1.0 (10)\%$ for light quark and gluon jets~\cite{BTV-16-002}.

\section{Event selection}
\label{sec:selection}

The goal of the event selection is to retain \ttH, \ttlnu, \ttll, \tllq, and \tHq events while excluding as many contributions from background processes as possible.
The analysis is split into categories with two same-sign leptons ($2\ell\ss$), three leptons ($3\ell$), and four leptons ($4\ell$), where $\ell$ refers to either a \Pe or \Pgm.
Events must also have a minimum number of jets, as well as  \PQb jets, with additional requirements that depend on the lepton flavor and multiplicity.
Single- and double-lepton triggers are used to collect events containing two leptons.
For events with three or more leptons, a combination of single-, double-, and triple-lepton triggers are used.

All events are required to have at least two leptons passing the tight selection.
Events where a pair of tight leptons with an invariant mass less than $12\GeV$ is found are rejected, to avoid backgrounds from light resonances.
In addition, events must have at least two jets with $\pt > 30\GeV$ to be reconstructed in the pseudorapidity ($\eta$) region, $\abs{\eta} < 2.4$.
One or more of the jets also need to pass the medium working point of the DeepCSV  \PQb tagging algorithm, as at least one top quark whose decay produces a bottom quark jet is present in all signal events.
No explicit identification requirements are placed on \Pgt leptons, which are allowed to enter the event selection via their decay products.

\subsection{\texorpdfstring{$2\ell \ss$}{Lg} category}
\label{sec:eventSelection_2lss}

The $2\ell \ss$ category primarily targets \ttH and \ttlnu signal events in which the $ \PQt\cPaqt$ system decays semileptonically, with an additional, identically charged lepton coming from the decay of a \PW boson produced in association with a top quark pair (in the case of \ttlnu), or coming from the decay of a \PW boson or a \Pgt lepton produced in the decay of the Higgs boson (in the case of \ttH).
In addition, the \ttll process may contribute with events in which there is at least one lepton that is not reconstructed or does not pass the selection.
Selected events are required to contain exactly two same-sign leptons passing the tight criteria, with the lepton of higher (lower) $\pt$ required to have $\pt > 25$ ($15$)\GeV.
Residual \ttbarJets background contributions are suppressed by requiring that the charge of all electrons and muons, which pass the relaxed object selection criteria, is well measured.
Electrons are required to pass two conditions which test the consistency between the independent measurements of the electron charge obtained from the position of the ECAL cluster and from its track, while muons must satisfy the condition that the estimated uncertainty on the $\pt$ of the muon track is below $20\%$.
The events are required to contain at least four jets with $\pt > 30\GeV$ and $\abs{\eta} < 2.4$.
At least two of these jets must be  \PQb jets, one of which must satisfy the medium working point of the DeepCSV  \PQb tagging algorithm, while the second is allowed to pass the loose working point.
Events containing more than two tight leptons are vetoed to avoid overlap with the $3\ell$ and $4\ell$ categories.

\subsection{\texorpdfstring{$3\ell$}{Lg} category}
\label{sec:eventSelection_3l}

The $3\ell$ category selects \ttlnu events in which all three \PW bosons decay leptonically; \ttll events in which the $ \PQt\cPaqt$ system decays semileptonically and the \cPZ boson decays to two charged leptons; \tllq events in which the top quark decays leptonically and the \cPZ decays to two charged leptons; and \ttH events in which the $\PH$ decays to \PW, \cPZ or \Pgt particles, at least one of which then decays leptonically (with one or more charged leptons also coming from the decay of the $ \PQt\cPaqt$ system).
Selected events are required to contain exactly three charged leptons passing the tight object selection criteria.
The three charged leptons are required to have $\pt > 25$, $15$, and $10\GeV$.
In the case that the third lepton is an electron, the requirement on it is instead $\pt > 15\GeV$ in order to stay above the trigger thresholds and keep the contributions from nonprompt electrons under control.
Two or more jets with $\pt > 30\GeV$ and $\abs{\eta} < 2.4$ are required, at least one of which must satisfy the medium working point of the DeepCSV  \PQb tag algorithm.
Two subcategories are defined according to whether a second jet passing the medium  \PQb tag is present.
This avoids incorrectly promoting \tllq events into the subcategory requiring two  \PQb jets, due to misidentification.

\subsection{\texorpdfstring{$4\ell$}{Lg} category}
\label{sec:eventSelection_4l}

The $4\ell$ category targets \ttll events in which all the \PW and \cPZ bosons decay leptonically, and \ttH events in which the $\PH$ decays into a pair of \PW bosons, where all \PW bosons decay leptonically; \cPZ bosons, where at least one \cPZ decays leptonically; or leptonically decaying \Pgt leptons.
Events selected in this category are required to contain four or more leptons passing the tight object selection criteria and passing $\pt$ thresholds of $\pt > 25$, $15$, $10$, and $10\GeV$ for the lepton of highest, second, third, and fourth highest \pt, respectively. 
In the case of electrons, the requirement on the third or fourth lepton is instead $\pt > 15\GeV$ for the same reasons as in the $3\ell$ category.
Two or more jets with $\pt > 30\GeV$ and $\abs{\eta} < 2.4$ are required.
As in the $2\ell \ss$ category, at least two of these jets must be  \PQb jets, one of which must satisfy the medium working point of the DeepCSV  \PQb tagging algorithm, while the second is allowed to pass the loose working point.

\subsection{Separation into subcategories}
\label{sec:eventSelection_subc}

Events in the $2\ell \ss$, $3\ell$, and $4\ell$ categories are further separated based on a number of criteria, as summarized in Table~\ref{tab:Categories}.
In the $2\ell \ss$ category, events are further separated based on lepton charge.  This allows us to take advantage of the fact that the $\ttW^{+}$ cross section is roughly a factor of 2 larger than that of the $\ttW^{-}$ cross section, so it is enhanced in $2\ell \ss$ events where both leptons are positively charged.
In the $3\ell$ category, we consider separately events which contain a same-flavor, oppositely charged pair of leptons with an invariant mass that falls within $10\GeV$ of the $m_{\cPZ}$, which primarily serves to create a region where the contribution from (on-shell) \ttZ is enhanced.  For $3\ell$ events that do not fall within this region, a classification based on the sum of lepton charges is used, considering events where the charge sum is positive separately from those where it is negative.  This again exploits the difference in cross section between $\ttW^{+}$ and $\ttW^{-}$.  In all $3\ell$ events, a classification is done based on whether the event contains exactly one jet passing the medium DeepCSV working point, or two or more jets passing the medium DeepCSV working point; the contribution from \tllq is enhanced in the former case.  For the $4\ell$ category, events are not split based on the invariant dilepton mass due to a small population of events in those bins.

\begin{table}[!t]
    \caption{Requirements for the different event categories.  Requirements separated by commas indicate a division into subcategories.  The  \PQb jet requirement on individual jets varies based on the lepton category, as described in the text.}
    \label{tab:Categories}
    \resizebox{\linewidth}{!}{
    \begin{tabular}{l c c c c}
        \hline
        Selection              & 2$\ell$\ss                                  & \multicolumn{2}{c}{$3\ell$} & $\ge$4$\ell$      \\ \hline
        Leptons                & Exactly 2 leptons                             & \multicolumn{2}{c}{Exactly 3 leptons}         & $\ge$4 leptons      \\
        Charge requirements    & $\sum_{\ell} q <0, \sum_{\ell} q > 0$ & $\sum_{\ell} q <0, \sum_{\ell} q > 0$ & \NA & \NA      \\
        Jet multiplicity       & 4, 5, 6, $\ge$7 jets                          & \multicolumn{1}{c}{2, 3, 4, $\ge$5 jets}      & 2, 3, 4, $\ge$5 jets &  2, 3, $\ge$4 jets   \\
        Number of $ \PQb$ jets & $\ge$2 $ \PQb$ jets                           & \multicolumn{1}{c}{1, $\ge$2 $ \PQb$ jets}    & 1, $\ge$2 $ \PQb$ jets & $\ge$2 $ \PQb$ jets \\
        Dilepton mass          & \NA                                             & $\abs{m_{\ell\ell} - m_{\cPZ}} > 10\GeV$ & $\abs{m_{\ell\ell} - m_{\cPZ}} \le 10\GeV$ & \NA \\
        \hline
    \end{tabular}}
\end{table}

Events in all categories are then separated into jet-multiplicity bins, which are used to fit to data and place limits on EFT parameters, as described in Section~\ref{sec:fitting}.  The $2\ell \ss$ and $3\ell$ categories are split into four jet multiplicity bins, resulting in 8 and 24 subcategories respectively.  The $4\ell$ category is split into three jet bins, bringing the total to 35 distinct signal region subcategories.

\section{Background estimation}
\label{sec:backgrounds}

Contributions to the selected event yields arise from a variety of background processes, which can be categorized as reducible or irreducible and are either estimated from data control regions or modeled using simulation.
A background is considered as reducible in case one or more of the reconstructed electrons or muons passing the tight object selection are not due to genuine prompt leptons, as defined in Section~\ref{sec:reco}.  

In the $2\ell\ss$ categories, a reducible background arises from events containing lepton pairs of opposite charge, mostly from \ttbarJets production, in which the charge of one lepton is mismeasured.  Both background contributions are determined from data using the same methods as in Ref.~\cite{HIG-17-018}.  A summary of these procedures is given in Sections~\ref{sec:backgroundEstimation_fakes} and~\ref{sec:backgroundEstimation_flips}. 

The dominant irreducible background processes are diboson production and (to a lesser extent) triboson production and are estimated using simulations.  In the $3\ell$ and $4\ell$ categories in particular, diboson production is the dominant overall background (among both reducible and irreducible sources).

The production of a top quark pair in association with a photon represents an additional, minor source of background.  It is typically due to an asymmetric photon conversion into an electron-positron pair, where one of the produced particles carries most of the photon energy while the other one is not reconstructed.  Even though this is a reducible background source, its contribution is estimated using simulation, since the isolated leptons arising from asymmetric conversion are well described in simulation.

\subsection{Background from misidentified leptons}
\label{sec:backgroundEstimation_fakes}

The background from nonprompt leptons is estimated from control samples in data, applying the measured rate at which nonprompt leptons pass the lepton selection criteria.  This rate, referred to as the fake rate, is measured from a multijet sample dominated by nonprompt leptons.  The data analyzed in this region are collected with single lepton triggers, except at low muon \pt, where the presence of an additional jet with $\pt > 40\GeV$ is required in the trigger.  The misidentification rate is defined as the probability for a lepton candidate that passes a relaxed lepton cut to pass the tight lepton selection.  The misidentification rate is extracted separately for electrons and muons and is measured as a function of the $\pt$ and $\eta$ of the nonprompt lepton.  Contamination from prompt leptons in the multijet sample is subtracted based on expectations from simulation.

Once the misidentification rates have been determined, they are applied to a selection called the application region (AR), which is identical to the signal region except that one or more of the leptons are required to fail the tight selection but pass the relaxed lepton selection instead.  An estimate of the misidentified-lepton background in the SR is obtained by applying appropriately chosen weights to the events selected in the AR.  Further details on the nonprompt-lepton background estimation technique can be found in Ref.~\cite{HIG-17-018}.

\subsection{Background from lepton charge mismeasurement}
\label{sec:backgroundEstimation_flips}
The lepton charge mismeasurement background in the $2\ell\ss$ categories is dominated by \ttbar events with two prompt leptons in which the sign of either prompt lepton is mismeasured.
This background contribution is estimated from data, following a strategy similar to the one used for the estimation of the nonprompt background.
It is found to be relevant only for electrons.
The electron charge is defined using the majority method, which takes the charge to be the one shared by at least two of the three charge estimate methods~\cite{Khachatryan:2015hwa}.
An AR is defined for the $2\ell\ss$ categories, requiring that the two selected leptons are of opposite charge.
The events in the AR are then weighted using the probability to mismeasure the electron charge, derived in a $\cPZ/\Pggx \to \Pe\Pe$ sample and parameterized as a function of the electron $\pt$ and $\eta$.
The probability for mismeasuring the sign of electrons ranges from 0.02\% for electrons in the barrel to 0.2\% for electrons in the endcaps, after all the object selection criteria.
The sum of the charge mismeasurement probabilities of the two lepton candidates is used to determine the overall background.

\section{Signal extraction}
\label{sec:fitting}

As stated in Section~\ref{sec:eventSelection_subc}, the analysis is split into 35 statistically independent categories, based on the sum of lepton charges, number of  \PQb tagged jets, and jet multiplicity.  A likelihood function is built based on independent bins following Poisson statistics.  The event yields are a function of the WCs and a set of nuisance parameters (NPs) which contain the effects of systematic uncertainties (see Section~\ref{sec:systematics}).  The WCs are parameterized by the quadratic form for the cross section as explained in Section~\ref{sec:samples}.
In order to fit this parameterization to the data, we scan over all WCs.  The boundaries of the scans roughly correspond to values of the chosen operator that result in a five-fold increase in the SM cross section for at least one signal process.  At each point in the scan, the negative log profiled-likelihood is computed, and the best fit is evaluated as the WC which minimizes the negative log likelihood.  Confidence intervals of 1 and 2 standard deviations ($\sigma$) are calculated by finding where the negative log likelihood curve crosses twice the value of one and four respectively.  In principle, this scan can be performed in the 16D WC space.  However, fitting this hypersurface is time-consuming, and the results are difficult to interpret.  Instead, we perform the fit for a single WC in two scenarios: when the other 15 WCs are treated as unconstrained NPs (profiled); and when the other 15 WCs are fixed to their SM value of zero.

\section{Systematic uncertainties}
\label{sec:systematics}

There are two types of systematic effects considered: those that affect only the rates of signal or background processes, and those that affect both the rate and the shape which refers to changes in the relative expected yield of the jet and/or  \PQb jet bins.  In the latter case, the rate and shape effects are treated simultaneously so that they are considered fully correlated.  Each systematic uncertainty is correlated across all analysis bins by using a single NP per physical process.  The sources of systematic uncertainties considered are: the integrated luminosity; the jet energy scale (JES);  \PQb jet tagging scale factors; the theoretical cross section; the PDF shape variations; the renormalization ($\muR$) and factorization ($\muF$) scales; the PS; the parton matching; the additional radiation; the muon and electron ID, isolation, trigger efficiency; the pileup; the the misidentified-lepton rate estimate; and the charge misreconstruction estimate.

All variations listed are applied equally to the signal and background samples, and are treated as 100\% correlated across all samples.  Unless otherwise stated, all systematic uncertainties are considered independent and thus uncorrelated with each other.  

\begin{itemize}
\item \textit{Integrated luminosity} The measured uncertainty on the LHC integrated luminosity estimate is 2.3\%~\cite{LUM-17-004}.

\item \textit{Jet Energy Scale}
The JES is adjusted via scale factors to account for pileup, nonuniform detector response, and any residual differences between the data and simulation.  The resulting effect of the JES uncertainty on the yields is determined by shifting the jet energy correction up and down by 1$\sigma$ , and propagating the changes through the object ID and event selection.

\item \textit{ \PQb jet tagging scale factors}
In order to use the DeepCSV tagger for identifying  \PQb jets, scale factors are applied to the simulated samples to bring them in agreement with data~\cite{BTV-16-002}.  There are three types of systematic uncertainties associated with the use of these scale factors: the JES, the purity of the control samples used to derive the scale factors, and the size of all the samples used to derive the scale factors.
The JES dependence is calculated simultaneously with the JES systematic uncertainty.  The purity component is treated by assigning NPs for the yields from both the light (\Pg,  \PQu,  \PQd,  \PQs) and heavy (\cPqc,  \PQb) flavors.  The \cPqc jet tagging uncertainty is used to remove \cPqc jets potentially mistagged as  \PQb jets.  Finally, the statistical uncertainty of the samples used to derive the scale factors is taken into account with four NPs: two for the light-flavor (LF) case and two for the heavy-flavor (HF) case.  The two NPs for each case are: an NP for the overall tilt that would be consistent with the statistical uncertainties on the SFs, while the second NP controls distortions of a more complicated nature, where the upper and lower ends of the distribution change relative to the center.  These NPs account for discrepancies in the shape of the tagging discriminant distributions, which are consistent with the uncertainty of the scale factors.

\item \textit{Theoretical cross section} The expected yields for signal and background are derived from theoretical predictions of at least NLO accuracy.  There are associated uncertainties on the $\muR$ and $\muF$ scales of the process and the PDF.  Table~\ref{tab:xsUncertainty} summarizes these uncertainties.  For signal processes, this uncertainty is considered on the whole process (SM+EFT).  These uncertainties do not vary with the WCs, so the uncertainties are of the same magnitude as when fixing the model to the SM component alone.

    \begin{table}[!tb]
      \centering
      \topcaption{Cross section (rate) uncertainties used for the fit.  Each column in the table is an independent source of uncertainty.  Uncertainties in the same column for different processes (different rows) are fully correlated.}
      \begin{tabular}{lrr}
        \hline
        Process & $\muR$ and $\muF$ scales & PDF \\
        \hline
        \ttH           & $-$9.2\% $+$5.8\%    & $\pm$3.6\% \\
        \tllq          & $\pm$1\%           & $\pm$4\%   \\
        \ttlnu         & $-$12\% $+$13\%      & $\pm$2\%   \\
        \ttll          & $-$12\% $+$10\%      & $\pm$3\%   \\
        \tHq           & $-$8\% $+$6\%        & $\pm$3.7\% \\
        Diboson          & $\pm$2\%           & $\pm$2\%   \\
        Triboson         & $\pm$2.6\%         & $\pm$4.2\% \\
        \ttgamma+jets  & $\pm$10\%          & $\pm$5\%   \\
        \hline
      \end{tabular}
      \label{tab:xsUncertainty}
    \end{table}

\item \textit{PDF shape variations} The shape variation of the final fitting variable distributions due to the uncertainty on the PDF is estimated by reweighting the spectra according to 100 replica sets.  The total uncertainty is measured using the standard PDF4LHC~\cite{Butterworth:2015oua} recommendation.

\item \textit{Renormalization and factorization scales} Uncertainties due to the $\mu_{\mathrm{R}}$ and $\mu_{\mathrm{F}}$ scales in the \ttbar{} ME generator are modeled by varying the scales independently by a factor of $1/2$ or 2 and propagating the changes to the final fitting variable distribution in the fit.  An uncertainty envelope is then calculated from these two systematic uncertainties.  This is accommodated via weights obtained directly from the generator information.  Since the normalization uncertainties of the ME generators are covered by the cross section uncertainties (listed in Table~\ref{tab:xsUncertainty}), only the impact on the kinematic shape of the process in question are considered.  These shape effects primarily enter as changes in the acceptance and efficiency for events to fall into a particular event selection category.  The bounds of the envelope are determined by taking the maximum of the $\mu_{\mathrm{F}}$ uncertainties, the $\mu_{\mathrm{R}}$ uncertainties, and their sum.  The $\mu_{\mathrm{R}}$ and $\mu_{\mathrm{F}}$ effects on the WCs range between 1 and 5\% depending on the bin.

\item \textit{Parton shower} The uncertainty in the PS simulation is estimated by varying the $\mu_{\mathrm{R}}$ for initial- and final-state radiation (ISR/FSR) up and down in \PYTHIA by multiplying/dividing the scale by a factor of 2 for ISR and $\sqrt{2}$ for FSR.  A dedicated SM sample (produced without EFT effects) is used to determine this systematic variation; the values obtained through this study are then applied to the full analysis samples.

\item \textit{Parton matching} This uncertainty only applies to \ttH, \ttW, and \ttZ processes, since matching is only performed for processes that include an extra parton.  Determined by varying the matching scale value between the extra partons generated in \MGvATNLO and jets produced in \PYTHIA, this uncertainty is computed bin-by-bin.  A dedicated SM sample (produced without any EFT effects) was used to perform this study.  The nominal scale is 19\GeV, and is shifted up to 25\GeV and down to 15\GeV.

\item \textit{Additional radiation} Since an extra final-state parton was not included in the LO single top quark processes (\tllq and \tHq), they are not expected to be as sensitive to varying the WCs as the LO \ttH, \ttW, and \ttZ samples.  A comparison of the LO \tllq sample to the NLO \tZq sample, reveals a discrepancy in the event yield, which is not covered by the existing systematic uncertainties.  We therefore introduced a new systematic uncertainty specifically for the \tllq sample to cover this disagreement.  The same systematic uncertainty is applied to \tHq, since the uncertainty accounts for the fact that \MGvATNLO cannot handle the matching for these extra partons for any $t$-channel process.  These uncertainties are typically around 20\%, but can reach as high as 80\% for the high jet multiplicity bins with few events.

\item \textit{Muon and electron ID and isolation} Scale factors are used to correct the tracking efficiency, electron and muon ID efficiency, and isolation in the simulation to match that in data, which are derived with a ``tag-and-probe" method~\cite{Sirunyan:2018fpa,Khachatryan:2015hwa,Khachatryan:2010xn}.  The impacts of these quantities are estimated by varying the scale factors within their uncertainties.  The resulting systematic uncertainties are typically of the order of 1--2\% per lepton.

\item \textit{Trigger efficiency} The impact due to the trigger efficiency~\cite{Sirunyan:2018fpa} is estimated by varying the trigger scale factors within their uncertainties, which are in the range of 2--5\%. 

\item \textit{Pileup} Effects due to the uncertainty in the distribution of the number of pileup interactions are evaluated by varying the total inelastic $\Pp\Pp$ cross section used to predict the number of pileup interactions in the simulation by 4.6\% from its nominal value, which corresponds to a 1$\sigma$ variation~\cite{Aaboud:2016mmw}.  This effect typically ranges from less than 1--3\%.

\item \textit{Misidentified-lepton rate estimate} 
Several sources of systematic uncertainty are considered.  The measurement 
of the misidentified-lepton weights is affected by the small population in the measurement region, subtraction of prompt lepton contamination in this region, as well as the uncertainty in the background jet composition in this region 
(dominated by multijet background) and the AR (dominated by \ttbarJets background).  
The effect on the misidentified lepton rate due to the overall uncertainty of the misidentified leptons is taken into account by varying the entire map of misidentified lepton weights up or down by 1$\sigma$.
This is the largest source of uncertainty on the misidentified lepton rate, and amounts to approximately 25--30\%, depending on the jet multiplicity bin. 
In addition, the limited population in the AR of the misidentified lepton method have 
a significant effect on the estimate of the misidentified lepton rate and must be considered as a separate source of uncertainty.  This again varies with jet multiplicity bin, and amounts to approximately 10--30\%.

\item \textit{Charge misidentification probability} The yield of the misreconstructed background in the 2$\ell \ss$ categories is known with an uncertainty of 30\%, and is included as a rate systematic uncertainty.  The uncertainty due to the limited population in the corresponding AR is negligible and is not considered.

\end{itemize}

Table~\ref{tab:systSummary} summarizes the systematic uncertainties assessed in the signal and backgrounds, and how each systematic uncertainty is treated in the fit used to extract the amount of signal present in the data sample.  We note that it is possible for the statistical and systematic uncertainties to depend on the choice of the initial WC values (used to evaluate the quadratic fit parameters).  To examine this, simulations are generated at the boundaries of the measured 2$\sigma$ confidence interval, and the SM point (\ie, all WCs set to zero), and no difference is observed within the current level of precision.  A summary of the percentage effect (change in WC divided by the symmetrized confidence interval) for \ctW, \ctp, \cQlM, and \ctl is provided in the last four columns of Table~\ref{tab:systSummary} to illustrate the range of values we observe for each systematic variation.  This table is related to the change in the WCs do to a single NP---correlations among NPs are not taken into account---and is therefore a conservative estimate.

\begin{table}[!t]
  \topcaption{Summary for the systematic uncertainties.  Here ``shape" means that the systematic uncertainty causes a change in the relative expected yield of the jet and/or  \PQb jet bins.  Except where noted, each row in this table will be treated as a single, independent NP.  Impacts of various systematic variations on a subset of WCs are also quoted.  Percentages represent the change in a WC divided by the symmetrized 2$\sigma$ confidence interval.  A value of 100\% indicates the particular systematic variation adds an uncertainty equal to the WC interval.  The percentages for the  \PQb and \cPqc jet tags are the sum of all their respective subcategories.}
  \label{tab:systSummary}
  \centering
  \begin{tabular}{l l r r r r}
    \hline
    Source                              & Type & \ctW & \ctp & \cQlM & \ctl \\
    \hline
    Integrated luminosity               & rate & 6\% & 2\% & 1\% & $<$1\% \\
    JES                                 & rate+shape & 6\% & 2\% & 1\% & $<$1\% \\
    \multicolumn{2}{l}{ \PQb jet tag} & 1\% & 5\% & 8\% & $<$1\% \\[\cmsTabSkip]
     \PQb jet tag HF fraction          & rate+shape \\
     \PQb jet tag HF stats (linear)    & rate+shape \\
     \PQb jet tag HF stats (quadratic) & rate+shape \\
     \PQb jet tag LF fraction          & rate+shape \\
     \PQb jet tag LF stats (linear)    & rate+shape \\
     \PQb jet tag LF stats (quadratic) & rate+shape \\[\cmsTabSkip]
    \multicolumn{2}{l}{\cPqc jet mistag} & $<$1\% & 12\% & 8\% & 2\% \\[\cmsTabSkip]
     \PQb jet tag charm (linear)       & rate+shape \\
     \PQb jet tag charm (quadratic)    & rate+shape \\[\cmsTabSkip]
    PDF ($\Pg\Pg$)                      & rate & 1\% & $<$1\% & $<$1\% & $<$1\% \\
    PDF ($\Pg\Pg_{\ttH}$)               & rate & $<$1\% & 1\% & $<$1\% & $<$1\% \\
    PDF (\qqbar)                      & rate & 1\% & $<$1\% & $<$1\% & $<$1\% \\
    PDF ($\cPq\Pg_{\tHq}$)              & rate & $<$1\% & $<$1\% & $<$1\% & $<$1\% \\
    $\muRF$ scale (\ttH)              & rate & 2\% & 5\% & $<$1\% & $<$1\% \\
    $\muRF$ scale (\ttgamma)          & rate & 1\% & 1\% & $<$1\% & $<$1\% \\
    $\muRF$ scale ($\ttbar \mathrm{V}$) & rate & 15\% & 4\% & 1\% & $<$1\% \\
    $\muRF$ scale (\tHq)              & rate & 1\% & 1\% & $<$1\% & $<$1\% \\
    $\muRF$ scale ($\mathrm{V}$)        & rate & $<$1\% & $<$1\% & $<$1\% & $<$1\% \\
    $\muRF$ scale ($\mathrm{VV}$)       & rate & $<$1\% & $<$1\% & $<$1\% & $<$1\% \\
    $\muRF$ scale ($\mathrm{VVV}$)      & rate & $<$1\% & $<$1\% & $<$1\% & $<$1\% \\
    PDF                                 & shape & 2\% & 1\% & $<$1\% & $<$1\% \\
    $\muRF$ scales                  & shape & $<$1\% & 6\% & 1\% & $<$1\% \\
    FSR                             & rate+shape & 1\% & 11\% & 7\% & 2\% \\
    ISR                             & rate+shape & $<$1\% & 8\% & 3\% & $<$1\% \\
    Parton matching                & rate+shape & 1\% & 10\% & 5\% & 1\% \\
    Additional radiation           & rate+shape & 11\% & 3\% & 1\% & $<$1\% \\
    Lepton ident./isol.                 & rate+shape & 4\% & 2\% & $<$1\% & $<$1\% \\
    Trigger efficiency                  & rate+shape & 2\% & 1\% & $<$1\% & 1\% \\
    Pileup                              & rate+shape & 1\% & 1\% & $<$1\% & $<$1\% \\
    Lepton misident.                      & rate+shape & 2\% & 70\% & 29\% & $<$1\% \\[\cmsTabSkip]
    Lepton misident.\stat                 & rate+shape \\[\cmsTabSkip]
    Charge misident.          & rate & 3\% & 2\% & $<$1\% & $<$1\% \\
    \hline
  \end{tabular}
\end{table}

\section{Results}
\label{sec:results}

\begin{figure}[!t]
	\centering
	\includegraphics[width=0.85\textwidth]{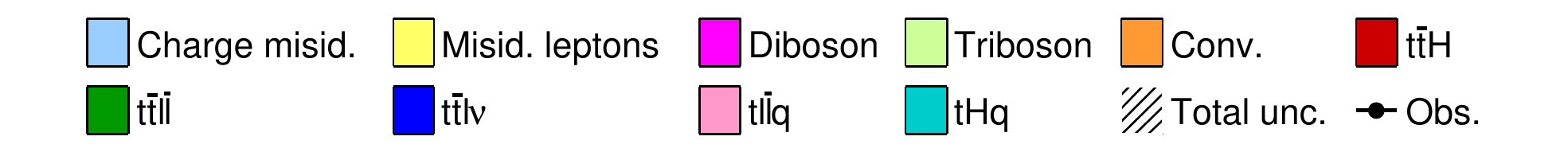} \\
	\includegraphics[width=0.49\textwidth]{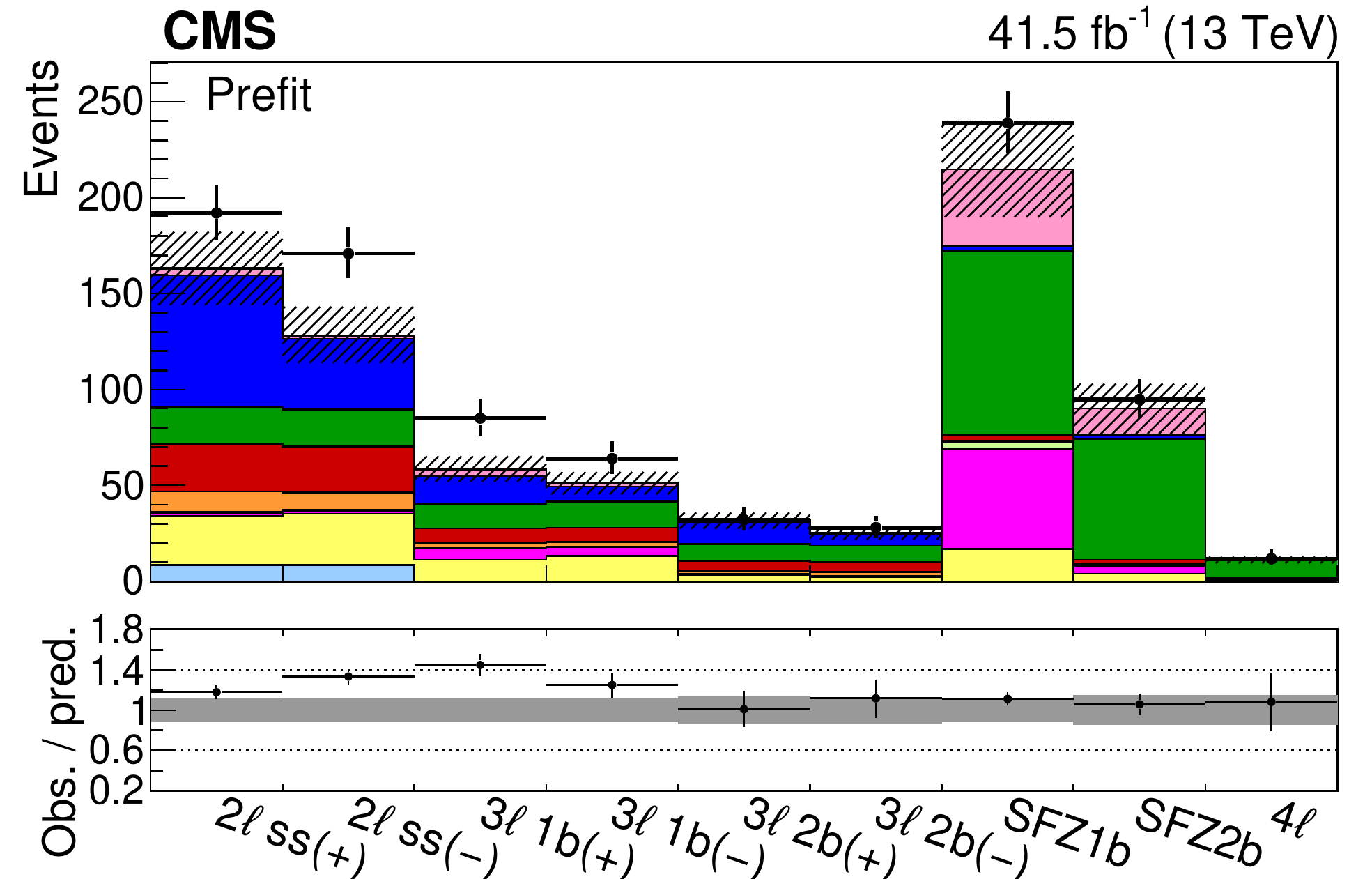}
	\includegraphics[width=0.49\textwidth]{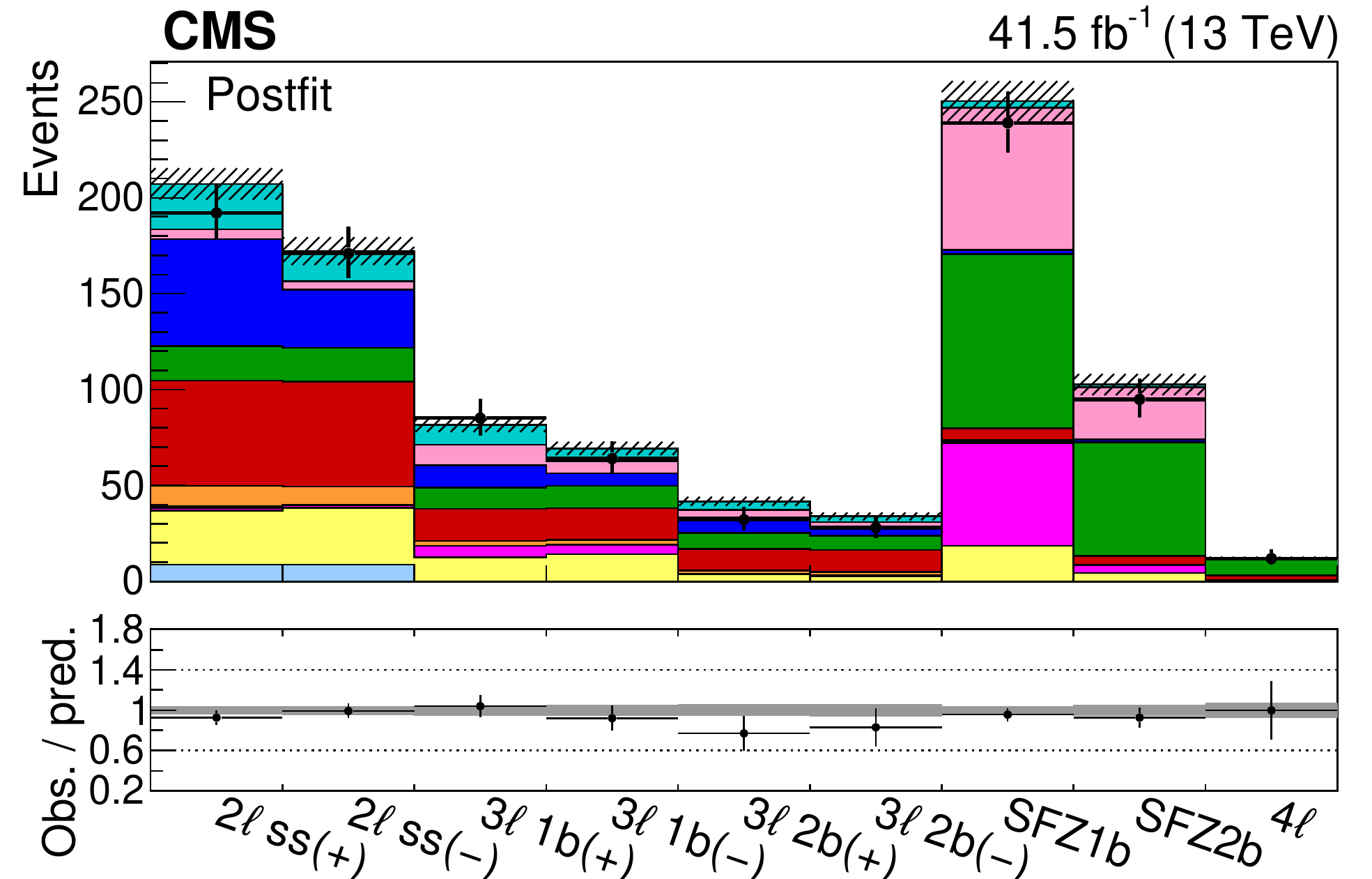}
	\caption{Expected yields prefit (left) and postfit (right). The postfit values of the WCs are obtained from performing the fit over all WCs simultaneously. ``Conv." refers to the photon conversion background, ``Charge misid." is the lepton charge mismeasurement background, and ``Misid. leptons" is the background from misidentified leptons.  The jet multiplicity bins have been combined here, however, the fit is performed using all 35 event categories outlined in Section~\ref{sec:eventSelection_subc}. The lower panel is the ratio of the observation over the prediction.}
	\label{fig:postfit-yields}
\end{figure}

The number of events selected in different categories is compared to the expected contributions of the signal processes and of the different background processes before (prefit) and after simultaneously fitting all 16 WCs, and the NPs, to minimize the negative log-likelihood (postfit) in Fig.~\ref{fig:postfit-yields}.  The prefit scenario corresponds to the SM where the values of the WCs are all assumed to be zero.  The simultaneous fit is equivalent to the 16 best fit points from the profiled fits: the profiled fit will always find the global minimum for each of the 15 profiled WCs.  The hatched region in the stack plot and the shaded region in the ratio plot show the sum of all systematic uncertainties.  The large increase in the \tHq event yields is a consequence of the low sensitivity to this particular process coupled with the fact that \tHq receives relatively large enhancements from the EFT operators considered.  The fit finds a combination of WCs that is able to enhance \tHq, which helps improve agreement in the $2\ell\ss$ and $3\ell$ (non-Z) categories, without spoiling the agreement elsewhere.  Despite the large increase, \tHq is still a smaller contribution than \ttH and \ttlnu in these categories.  There is also a large increase in the \tllq event yields, which we are also insensitive to.

\begin{figure}
\center
\includegraphics[scale=0.8]{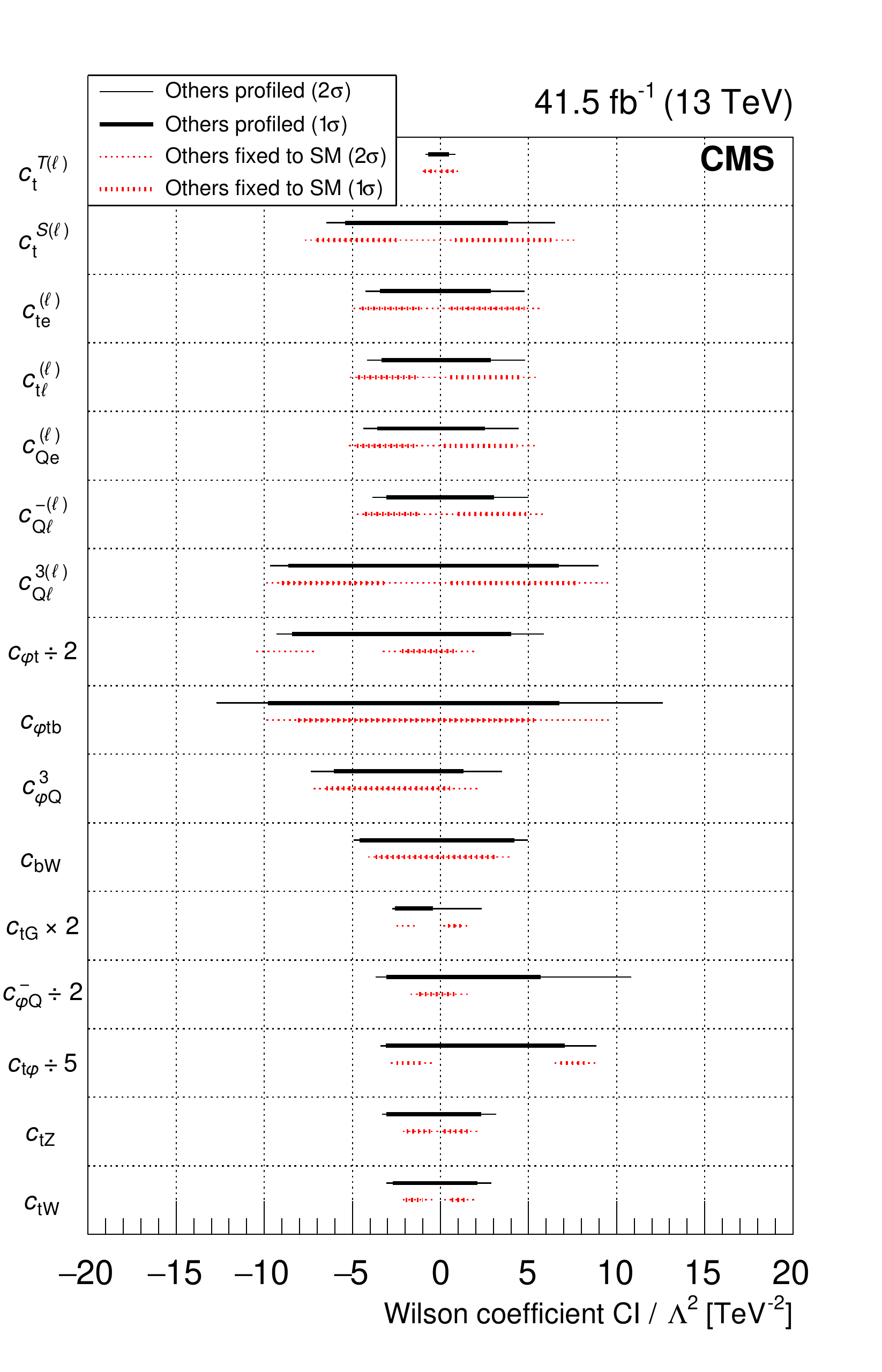}
\caption{Observed WC 1$\sigma$ (thick line) and 2$\sigma$ (thin line) confidence intervals (CIs). Solid lines correspond to the other WCs profiled, while dashed lines correspond to the other WCs fixed to the SM value of zero. In order to make the figure more readable, the \cpt interval is scaled by $1/2$, the \ctG interval is scaled by 2, the \cpQM interval is scaled by $1/2$, and the \ctp interval is scaled by $1/5$.}
\label{fig:SummaryPlot}
\end{figure}

Table~\ref{tab:Summary} shows the 2$\sigma$ confidence intervals for each WC.  Intervals are given for two scenarios. These results are displayed graphically in Fig.~\ref{fig:SummaryPlot}, along with their 1$\sigma$ confidence intervals (thicker lines).  The confidence intervals for a single WC (solid bars) are calculated while the other 15 WCs are profiled.  An alternative determination of the confidence interval for a single WC is performed by fixing the other 15 WCs to their SM values of zero (dashed bars).  For the profiled scenario, the confidence interval for all WCs includes the SM.  Occasionally, when fixing the other 15 WCs to zero the SM point falls just outside the 2$\sigma$ confidence interval (\eg, \ctW).  This is not surprising because in the cases where all WCs but one are fixed to zero, that single WC must account entirely for any deviation between observed data and expectation.  In contrast, in the profiled case, all 16 WCs can work together to accommodate any deviations, resulting in a best fit point that is closer to the SM, leading to the SM point falling inside the 2$\sigma$ confidence interval.  It is also sometimes possible for the profiled case to produce a more narrow 2$\sigma$ confidence interval, as can be seen for \ctlS, \cte, \ctl, \cQe, and \cQlM.  It is important to note that these five parameters each have disjoint nonzero 1$\sigma$ confidence interval when the other 15 parameters are frozen.  This will inherently broaden the profiled likelihood curve, resulting in a larger interval.  Note that as mentioned in Section~\ref{sec:signal_sample_generation}, the definition of \qq{}{ \PQu G}{} here includes an explicit factor of the strong coupling constant, which should be accounted for when comparing to results extracted based on other conventions.

\begin{table}
\begin{center}
\topcaption{The 2$\sigma$ confidence intervals on the WCs.  The intervals are found by scanning over a single WC while either treating the other 15 profiled, or fixing the other 15 to the SM value of zero.}
\begin{tabular}{ c c c }
\hline
WC$/\Lambda^{2}$[$\TeV^{-2}$] & 2$\sigma$ interval (others profiled) & 2$\sigma$ interval (others fixed to SM) \\
\hline
\ctW & [$-$3.08, 2.87] & [$-$2.15, $-$0.29]$\cup$[0.21, 1.96] \\
\ctZ & [$-$3.32, 3.15] & [$-$2.14, 2.19] \\
\ctp & [$-$16.98, 44.26] & [$-$14.12, $-$1.46]$\cup$[32.30, 44.48] \\
\cpQM & [$-$7.59, 21.65] & [$-$3.45, 3.33] \\
\ctG & [$-$1.38, 1.18] & [$-$1.26, $-$0.69]$\cup$[0.08, 0.79] \\
\cbW & [$-$4.95, 4.95] & [$-$4.12, 4.09] \\
\cpQa & [$-$7.37, 3.48] & [$-$7.21, 2.25] \\
\cptb & [$-$12.72, 12.63] & [$-$9.87, 9.67] \\
\cpt & [$-$18.62, 12.31] & [$-$20.91, $-$14.10]$\cup$[$-$6.52, 4.24] \\
\cQla & [$-$9.67, 8.97] & [$-$9.91, 9.50] \\
\cQlM & [$-$4.02, 4.99] & [$-$4.76, 5.83] \\
\cQe & [$-$4.38, 4.59] & [$-$5.20, 5.36] \\
\ctl & [$-$4.29, 4.82] & [$-$5.15, 5.51] \\
\cte & [$-$4.24, 4.86] & [$-$4.97, 5.80] \\
\ctlS & [$-$6.52, 6.52] & [$-$7.70, 7.70] \\
\ctlT & [$-$0.84, 0.84] & [$-$1.01, 1.01] \\
\hline
\end{tabular}
\label{tab:Summary}
\end{center}
\end{table}

\begin{figure}
\includegraphics[width=0.5\textwidth]{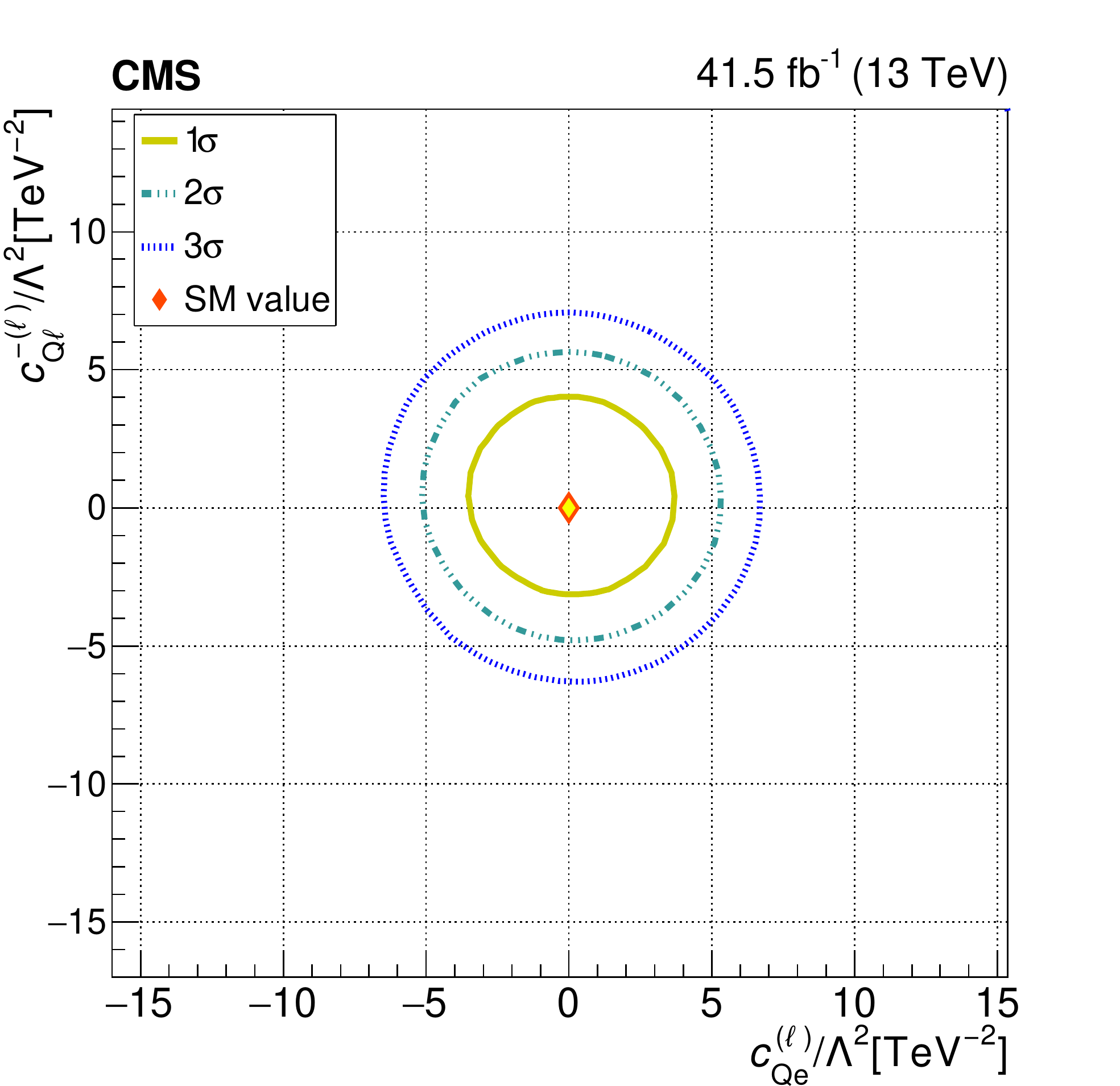}
\includegraphics[width=0.5\textwidth]{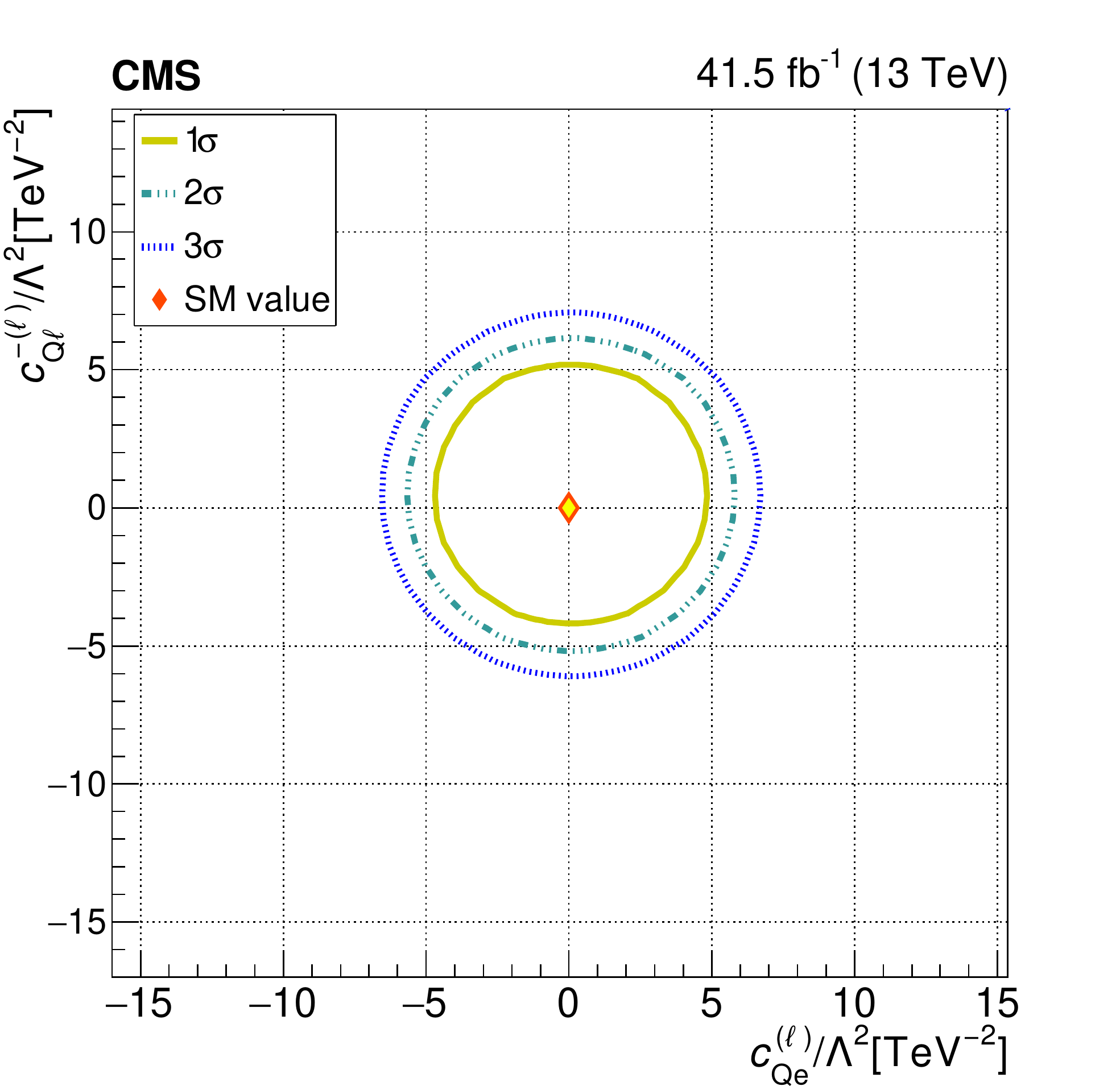}
\caption{The observed 1$\sigma$, 2$\sigma$, and 3$\sigma$ confidence contours of a 2D scan for \cQlM and \cQe with the other WCs profiled (left), and fixed to their SM values (right). Diamond markers are shown for the SM prediction.}
\label{fig:Contours_cQlmi_cQei}
\end{figure}

\begin{figure}
\includegraphics[width=0.5\textwidth]{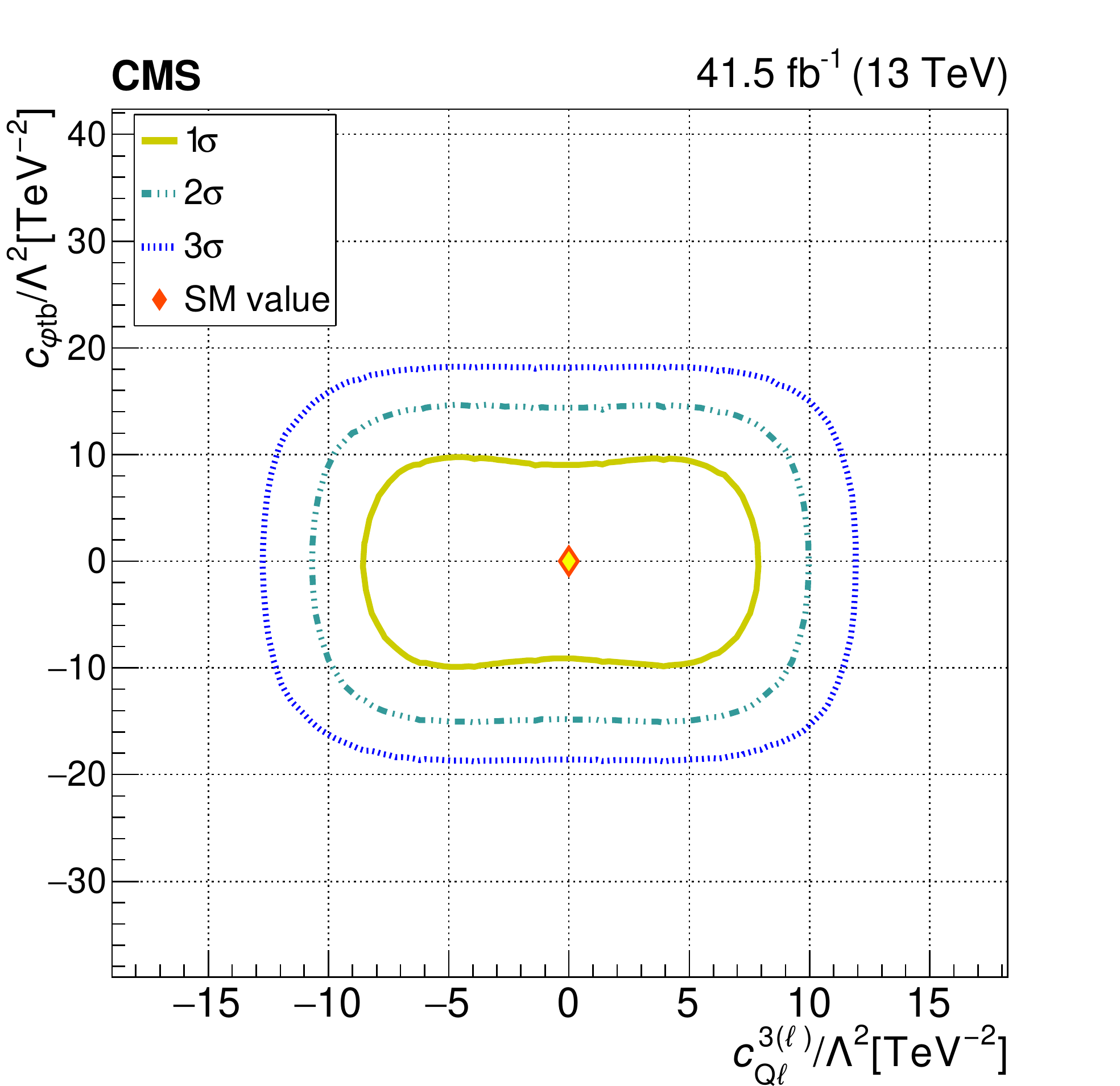}
\includegraphics[width=0.5\textwidth]{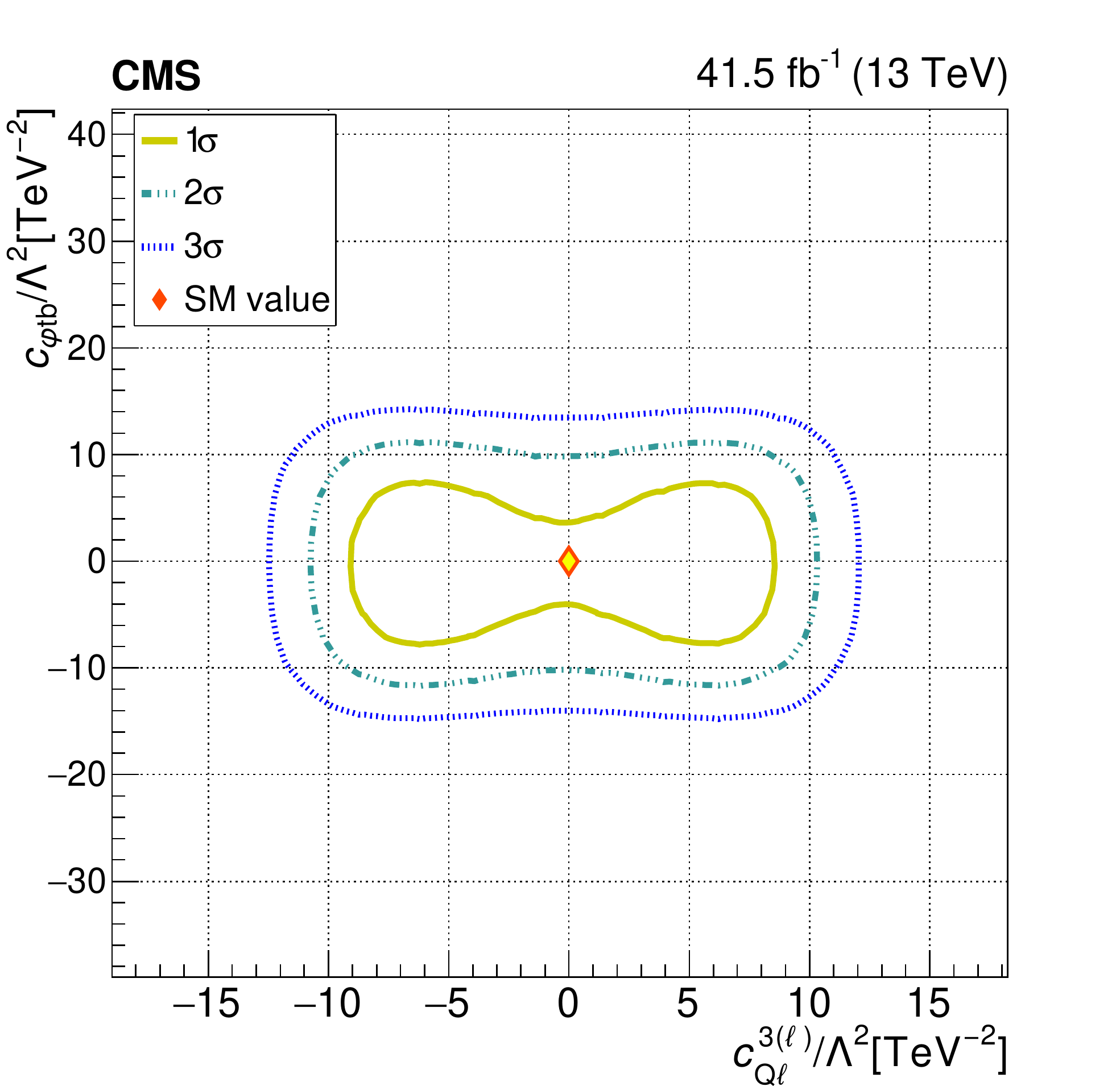}
\caption{The observed 1$\sigma$, 2$\sigma$, and 3$\sigma$ confidence contours of a 2D scan for \cptb and \cQla with the other WCs profiled (left), and fixed to their SM values (right). Diamond markers are shown for the SM prediction.}
\label{fig:Contours_cptb_cQl3i}
\end{figure}

\begin{figure}
\includegraphics[width=0.5\textwidth]{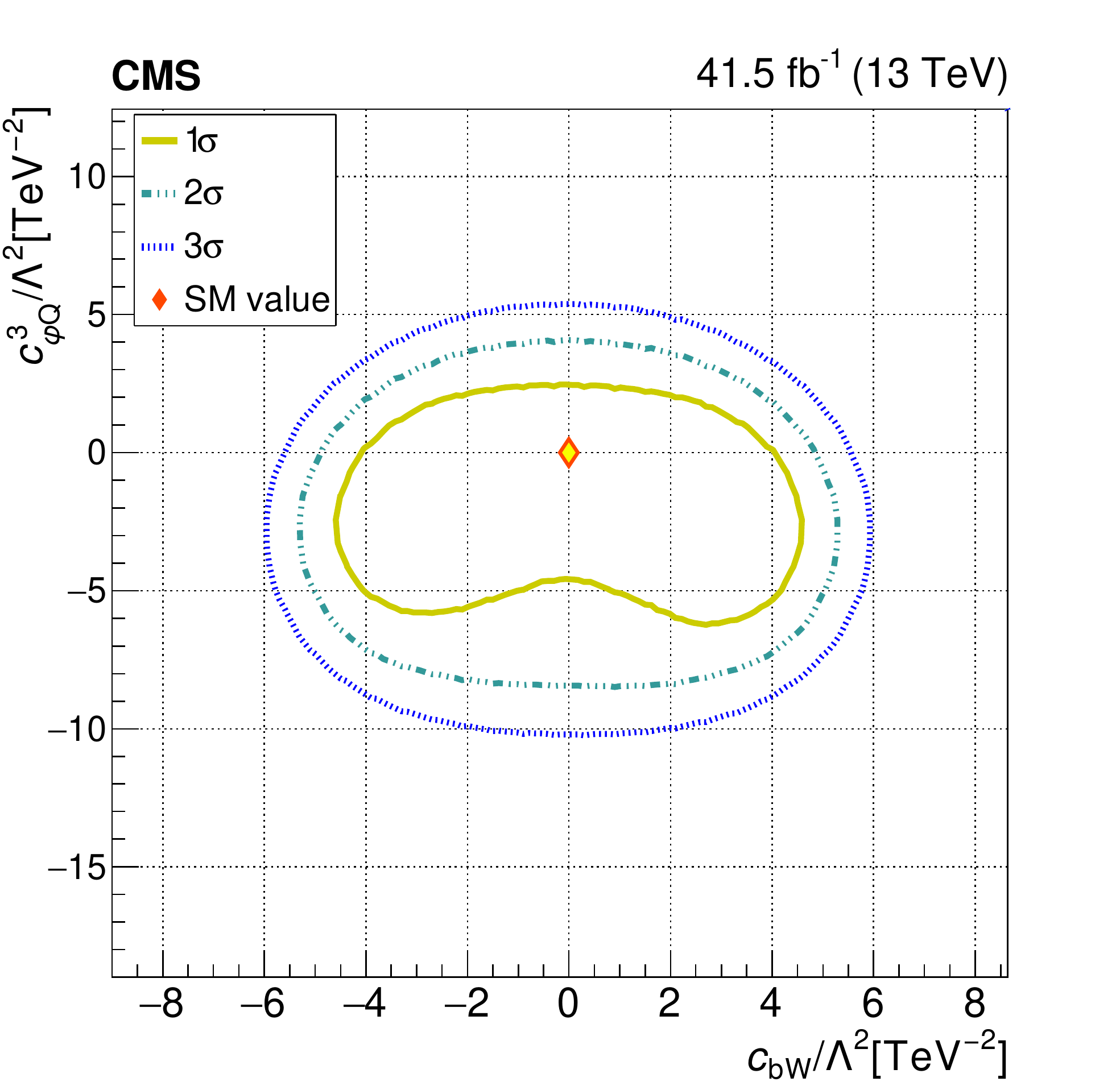}
\includegraphics[width=0.5\textwidth]{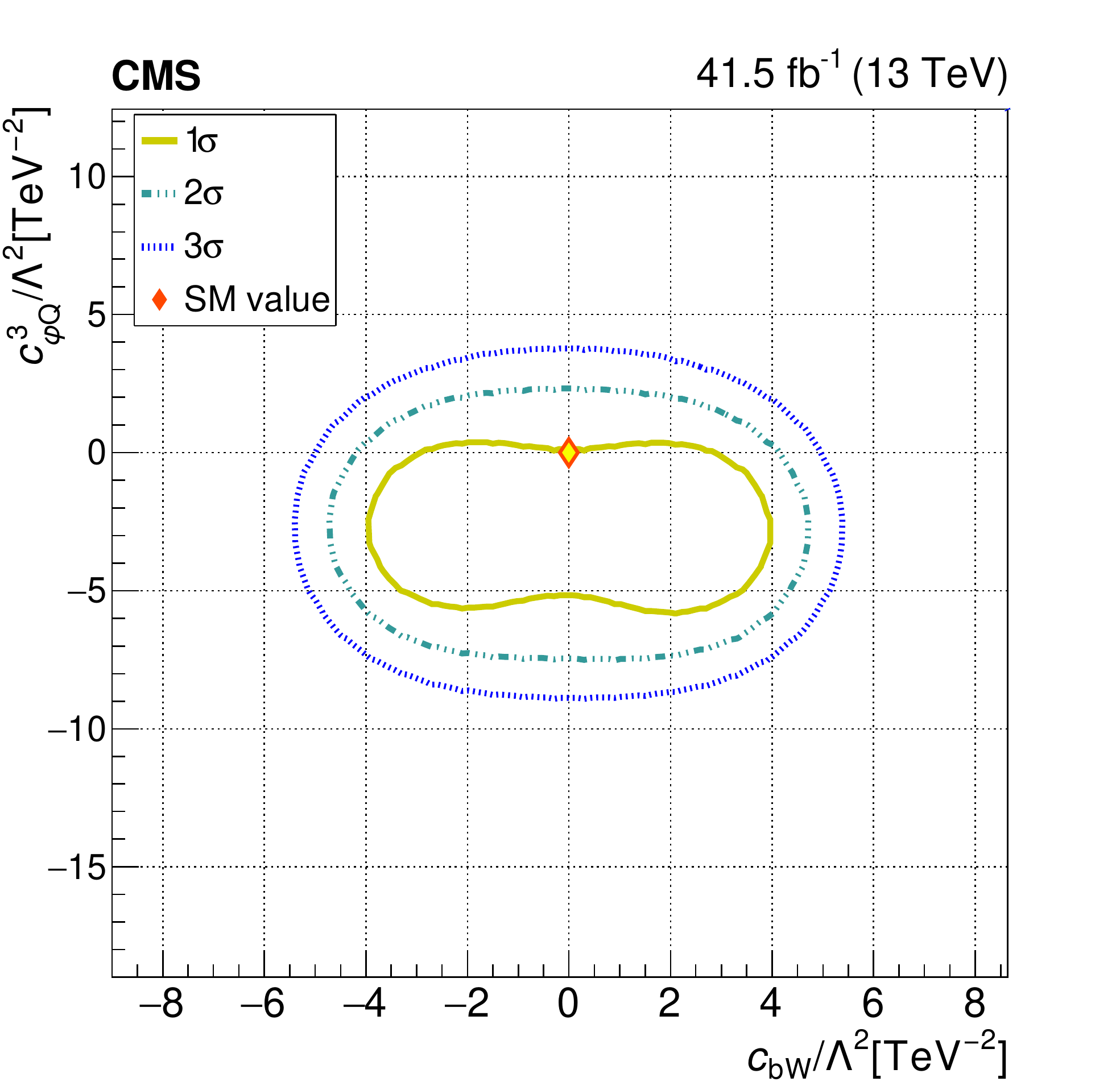}
\caption{The observed 1$\sigma$, 2$\sigma$, and 3$\sigma$ confidence contours of a 2D scan for \cpQa and \cbW with the other WCs profiled (left), and fixed to their SM values (right). Diamond markers are shown for the SM prediction.}
\label{fig:Contours_cpQ3_cbW}
\end{figure}

\begin{figure}
\includegraphics[width=0.5\textwidth]{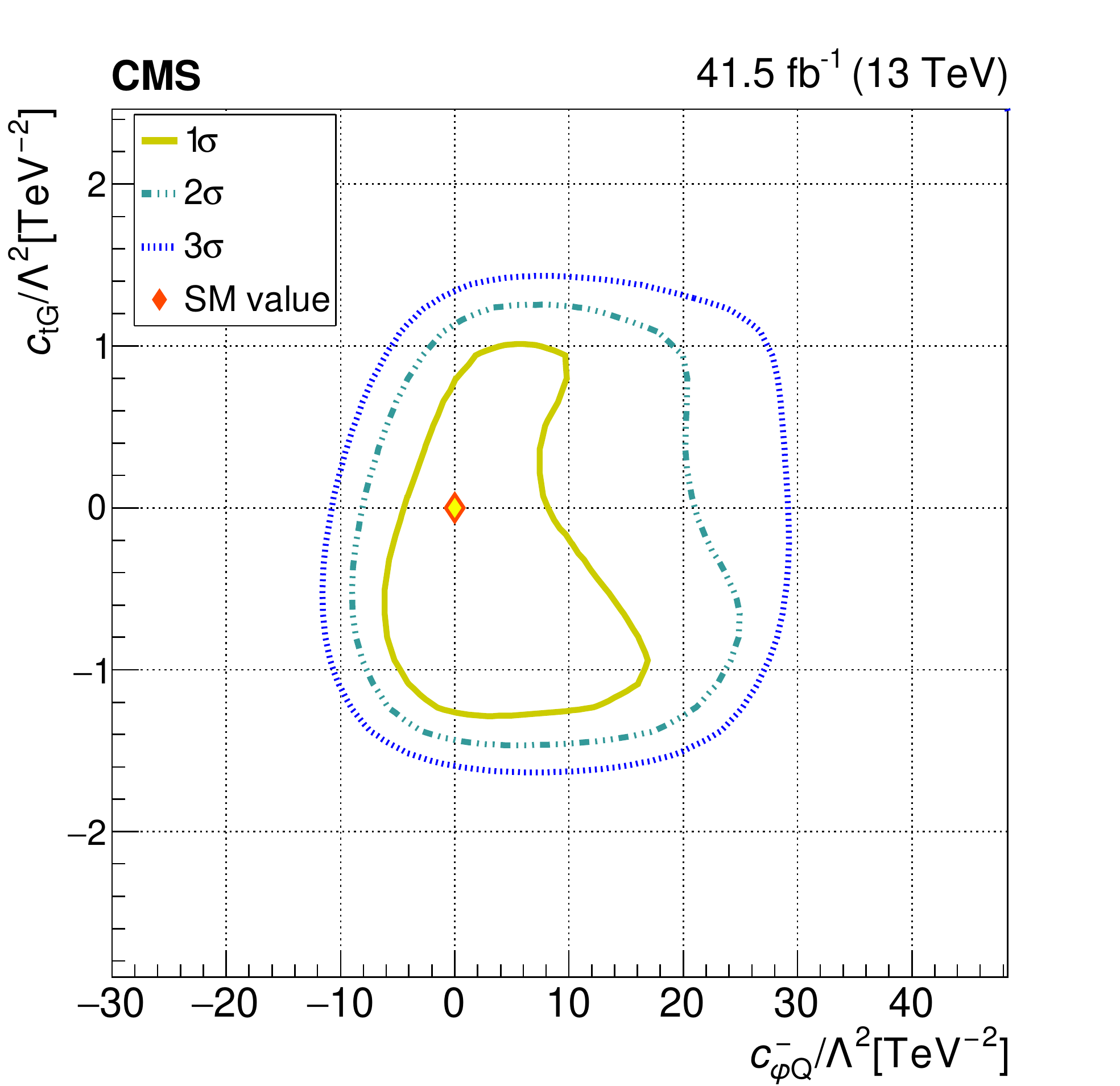}
\includegraphics[width=0.5\textwidth]{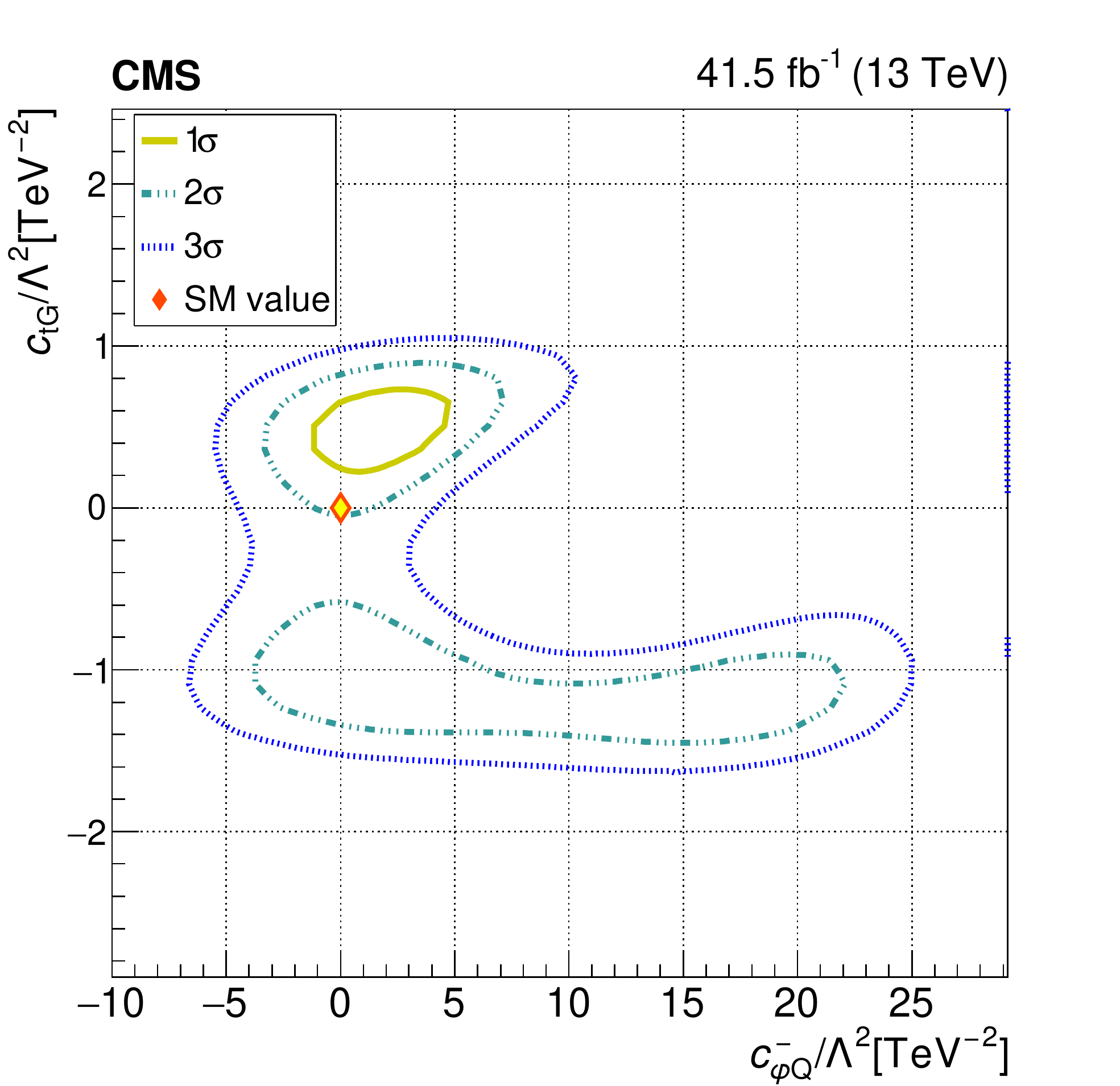}
\caption{The observed 1$\sigma$, 2$\sigma$, and 3$\sigma$ confidence contours of a 2D scan for \ctG and \cpQM with the other WCs profiled (left), and fixed to their SM values (right). Diamond markers are shown for the SM prediction.  The range on the right plot is modified to emphasize the 1$\sigma$ contour.}
\label{fig:Contours_ctG_cpQM}
\end{figure}

\begin{figure}
\includegraphics[width=0.5\textwidth]{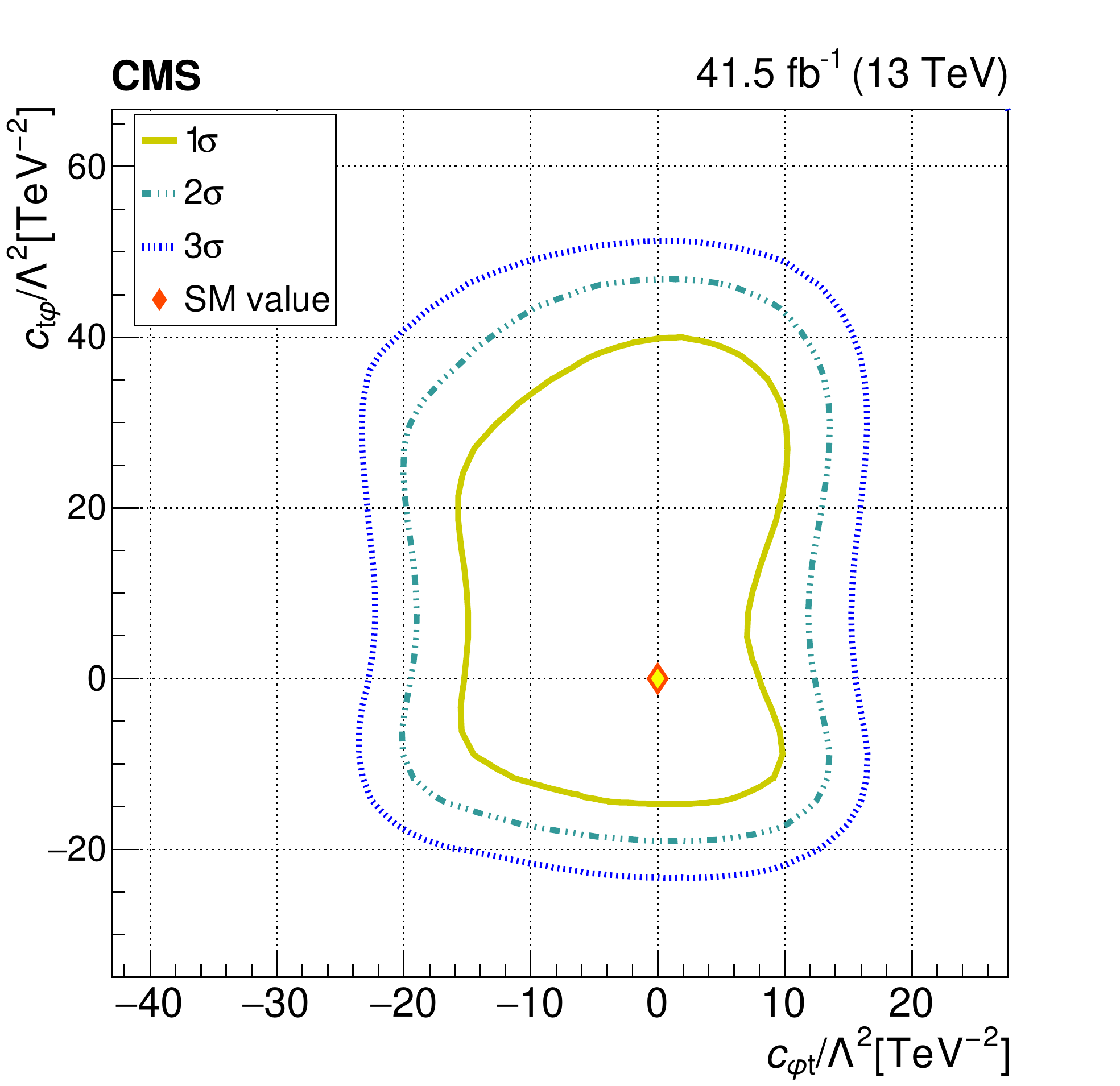}
\includegraphics[width=0.5\textwidth]{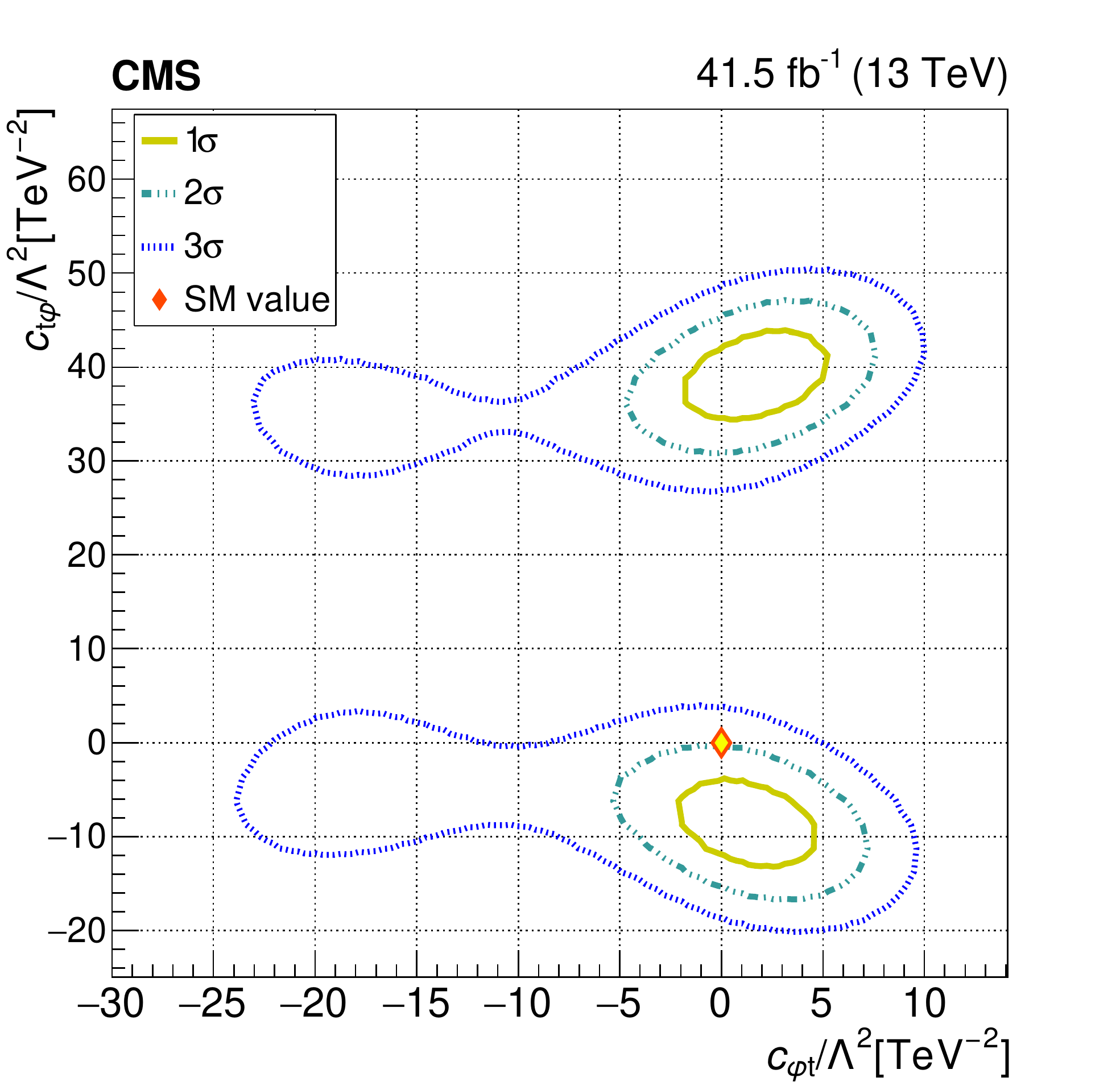}
\caption{The observed 1$\sigma$, 2$\sigma$, and 3$\sigma$ confidence contours of a 2D scan for \ctp and \cpt with the other WCs profiled (left), and fixed to their SM values (right). Diamond markers are shown for the SM prediction.  The range on the right plot is modified to emphasize the 1$\sigma$ contour.}
\label{fig:Contours_ctp_cpt}
\end{figure}

\begin{figure}
\includegraphics[width=0.5\textwidth]{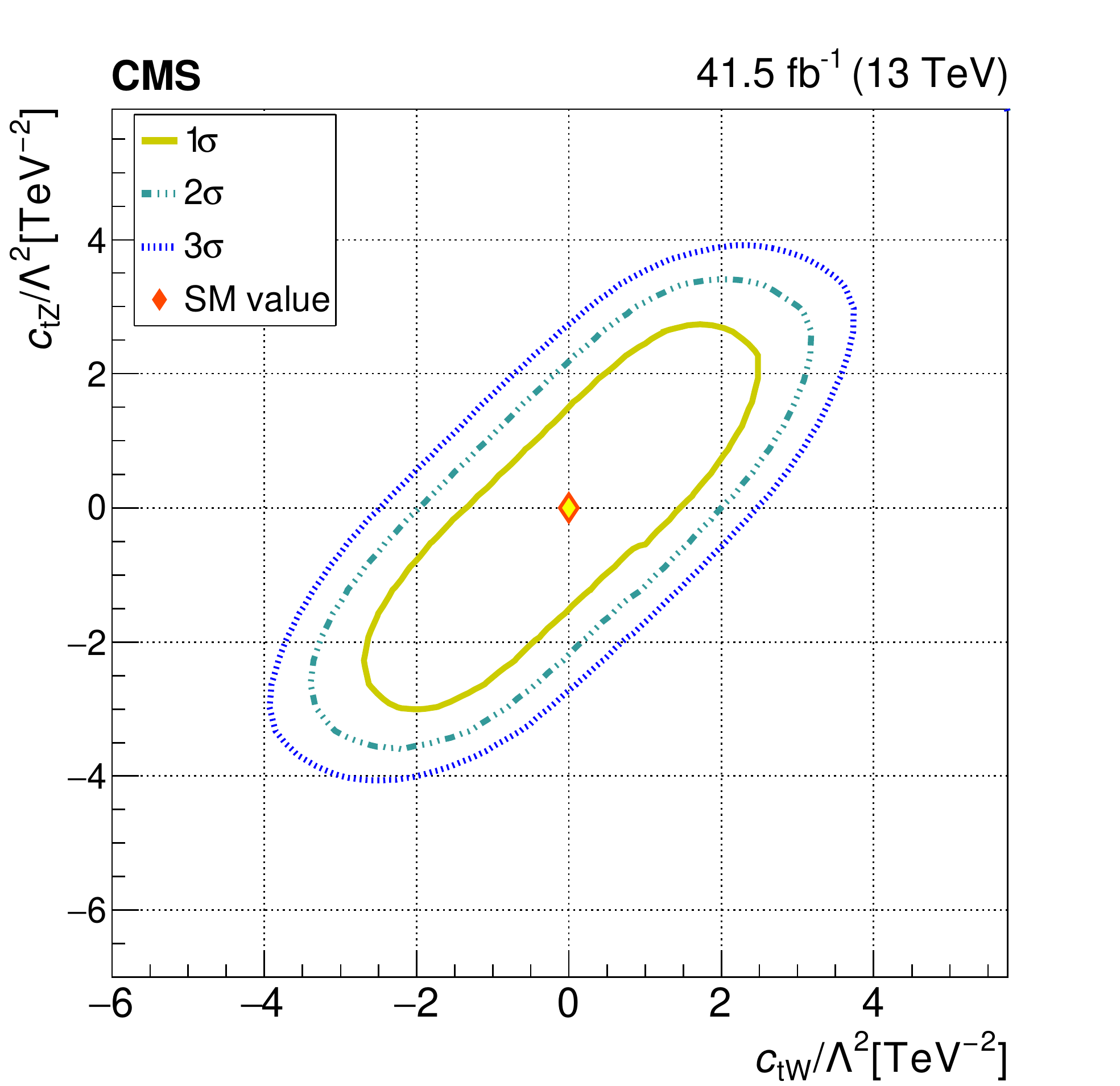}
\includegraphics[width=0.5\textwidth]{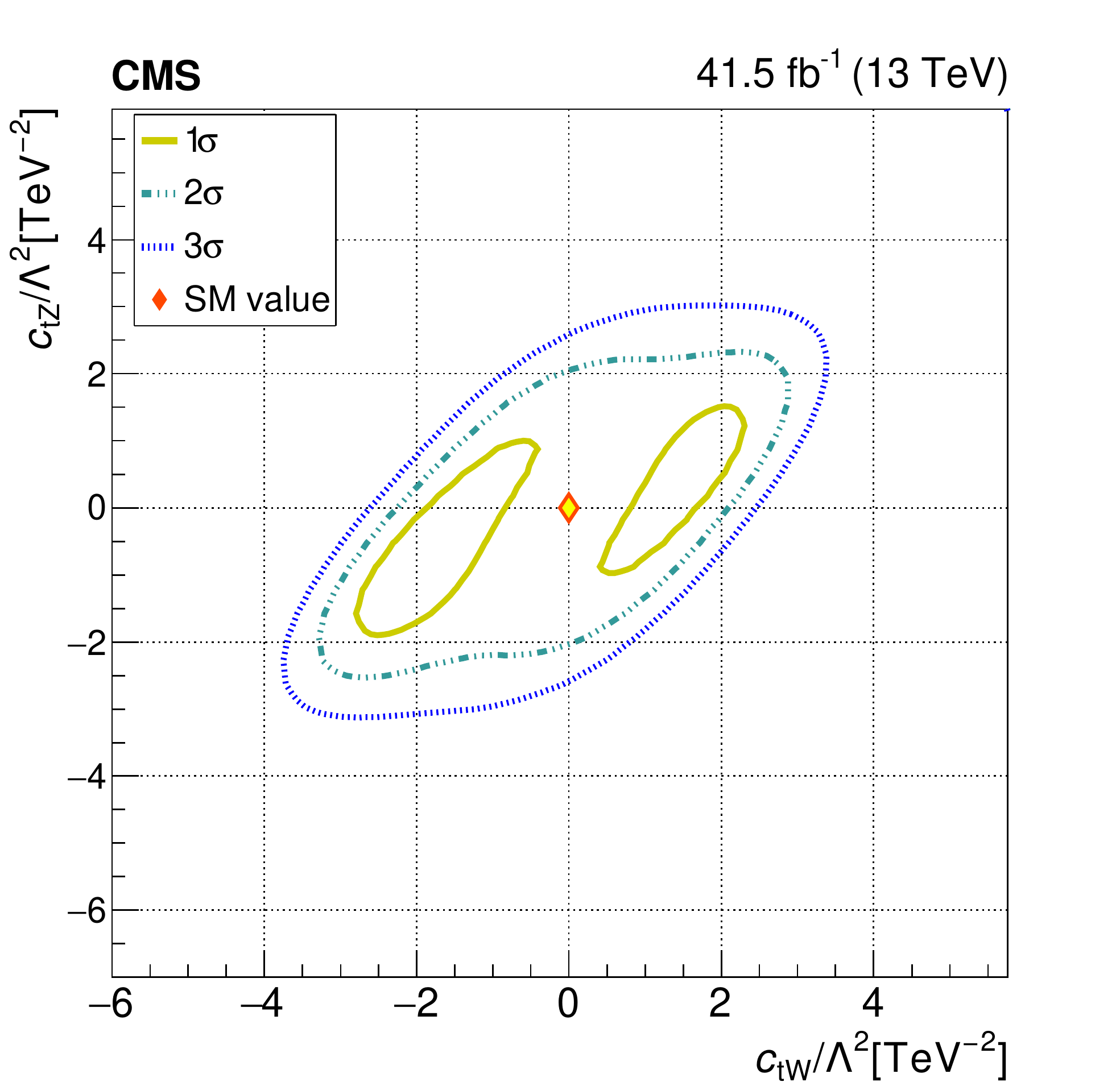}
\caption{The observed 1$\sigma$, 2$\sigma$, and 3$\sigma$ confidence contours of a 2D scan for \ctZ and \ctW with the other WCs profiled (left), and fixed to their SM values (right). Diamond markers are shown for the SM prediction.}
\label{fig:Contours_ctZ_ctW}
\end{figure}

Additional validation of the results was done with simultaneous scans in two WCs.  These are illustrated in Figs.~\ref{fig:Contours_cQlmi_cQei}--\ref{fig:Contours_ctZ_ctW}.  This subset of figures are the only ones which show clear signs of correlations between pairs of WCs.  The diamond marker corresponds to the SM theoretical prediction.  The contours correspond to the 1$\sigma$ (solid lines), 2$\sigma$ (thick dashed lines), and 3$\sigma$ (dash-dotted lines) confidence intervals.  Figure~\ref{fig:Contours_cQlmi_cQei} indicates minimal correlation between the WCs \cQlM and \cQe.  This is a common case for many coefficient pairs, especially when the other 14 are profiled.  The left panel of Fig.~\ref{fig:Contours_cptb_cQl3i} shows a more rectangular contour for \cptb and \cQla.  This indicates that these two WCs are not correlated, and the rectangular shape comes from a double minima in \cQla.  Comparing the left and right panels of Fig.~\ref{fig:Contours_cptb_cQl3i} shows \cQla has little preference over profiling the other 14 WCs versus fixing them to zero, while the \cptb range is cut almost in half when fixing the other 14 coefficients to zero.  Figure~\ref{fig:Contours_cpQ3_cbW} shows a similar behavior for \cpQa and \cbW.  The horizontal scales on the right panels of Figs.~\ref{fig:Contours_ctG_cpQM} and~\ref{fig:Contours_ctp_cpt} were modified to facilitate the display of the 1$\sigma$ confidence interval contour.  The right panel of Fig.~\ref{fig:Contours_ctZ_ctW} shows a preference in \ctW and \ctZ for nonzero values when the other 14 WCs are fixed to zero.  When the other 14 WCs are profiled (Fig.~\ref{fig:Contours_ctZ_ctW} left), the preference vanishes.  This is indicative of the complex interplay between all 16 WCs.  Exploring the 16D hypersurface provides a unique handle on the WCs, which was not previously utilized in single parameter analyses such as~\cite{Sirunyan:2017uzs, Sirunyan:2018ucr, Sirunyan:2019wka,CMS:2018jcg, CMS:2019too, Aad:2019pxo, Aaboud:2018nyl, Buckley:2015lku, Hartland:2019bjb, Brivio:2019ius}.  As a result, any differences compared to other analyses should not be surprising, and may indicate our robust technique more accurately captures the nuances of the EFT.

\begin{figure}[!t]
\centering
\includegraphics[width=0.85\textwidth]{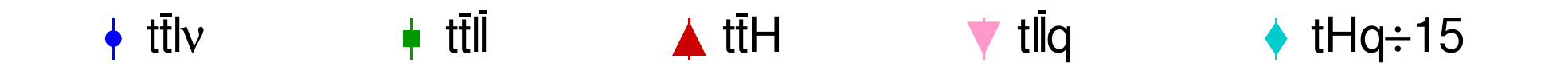} \\
\includegraphics[width=0.49\textwidth]{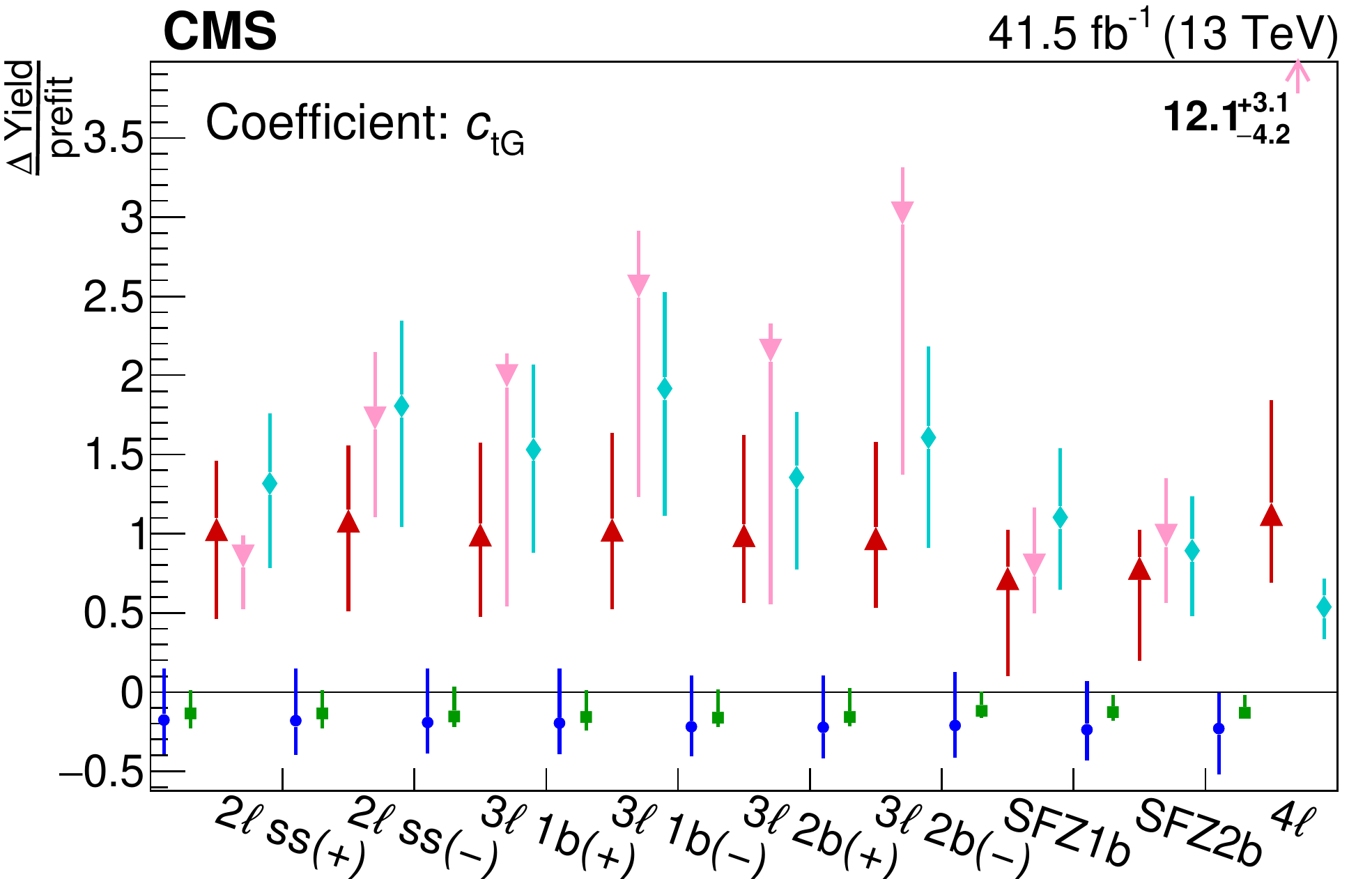}
\includegraphics[width=0.49\textwidth]{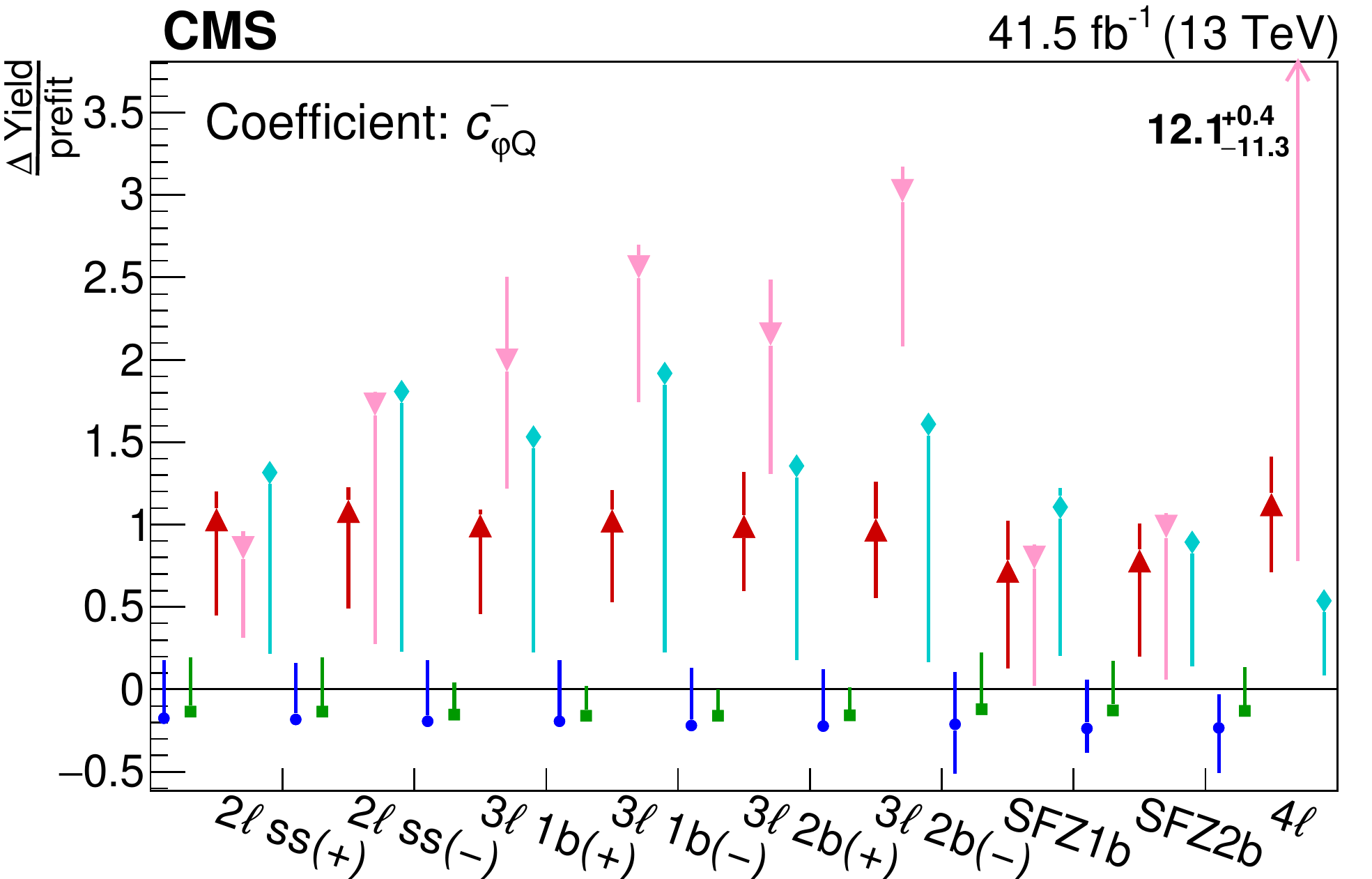} \\
\includegraphics[width=0.49\textwidth]{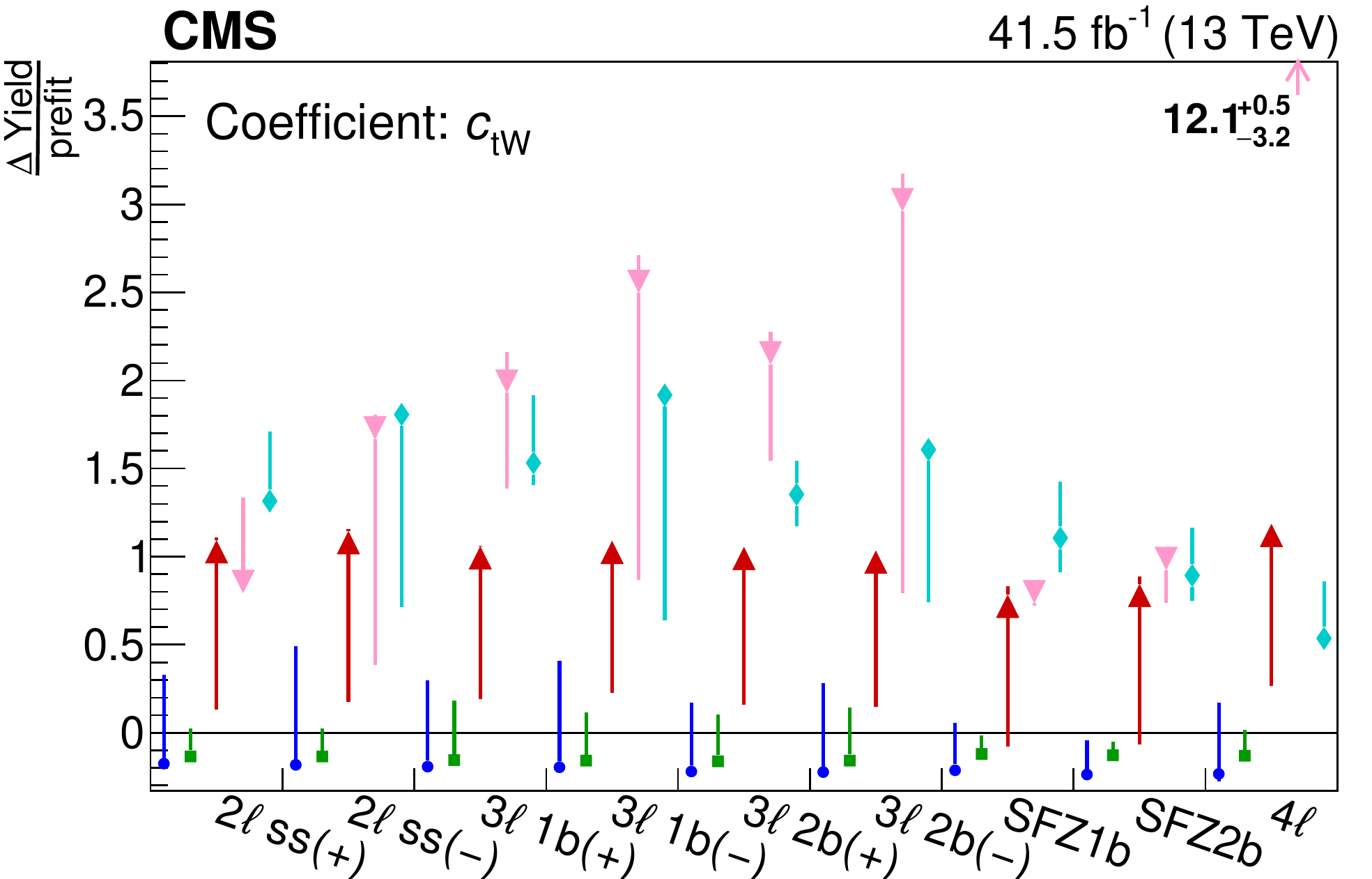}
\includegraphics[width=0.49\textwidth]{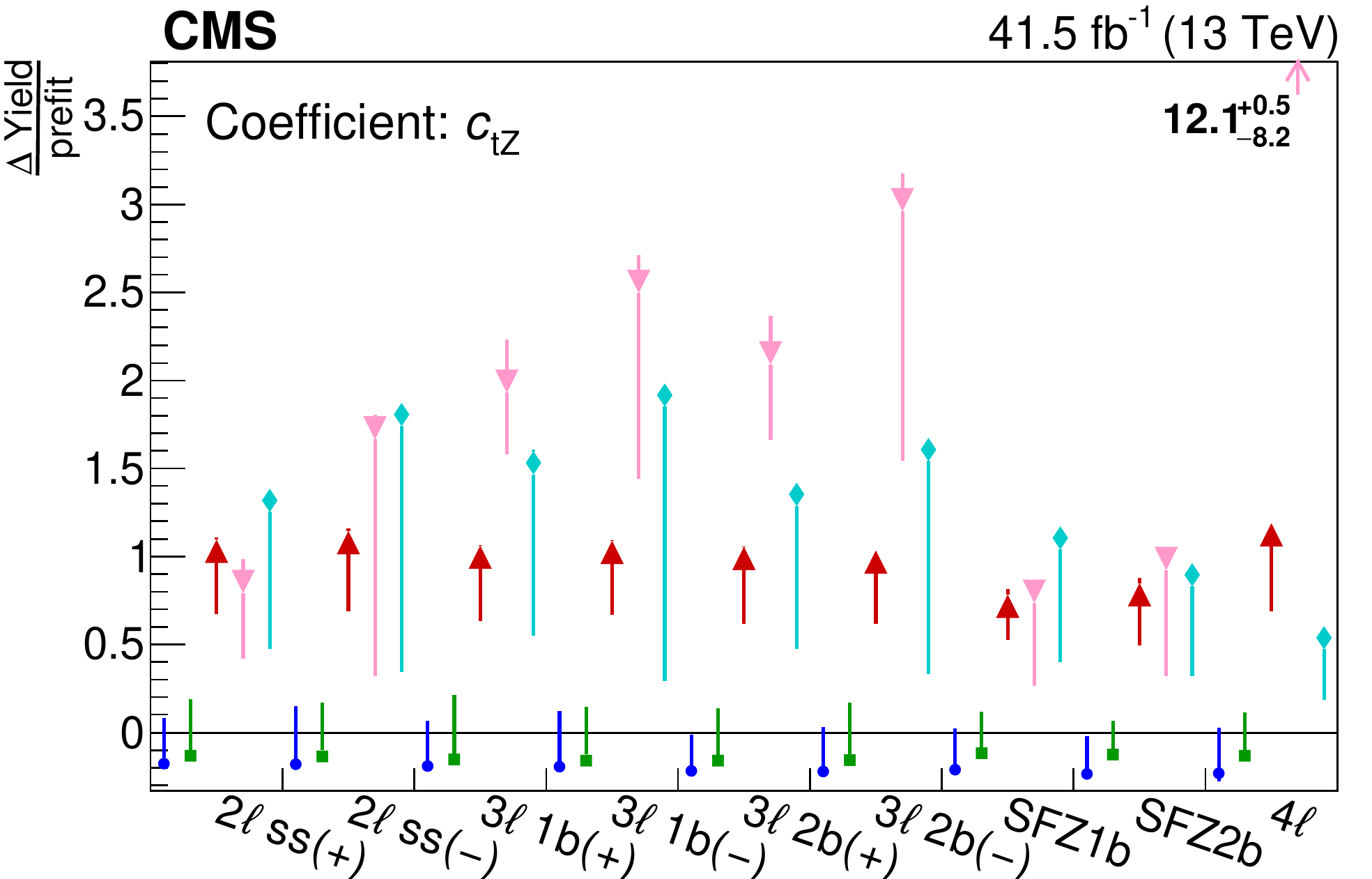} \\
\caption{Plots showing the relative change in the expected yield for the signal processes in each event category.  The ``$\Delta$Yield/prefit'' is the difference in expected yield before fit (prefit) and after fit (postfit), normalized to the prefit yield of the process in the corresponding category.  The vertical bars represent the maximum variation for a given WC within the corresponding 2$\sigma$ confidence interval.  The values in upper right of each plot are to indicate the variation for \tllq in the $4\ell$ category.}
\label{fig:yield-variations-all-categories}
\end{figure}

Figure~\ref{fig:yield-variations-all-categories} shows the fractional variation in expected yields for a given process and category after the fit and relative to the SM expectation.  The nominal points correspond to the best fit values obtained from the simultaneous fit of all WCs and are therefore the same in each plot.  The vertical bars represent the maximum variation in the expected yield within the corresponding 2$\sigma$ confidence interval for the given WC.  The variations are found by profiling all other WCs and NPs, that is by re-running the simultaneous fit for the other WCs and NPs, at each point in the scan for the WC of interest.  As the quadratic parameterization for a given WC need not be the same for each process or for each bin of a given process, the extrema of the vertical bars do not necessarily correspond to the same WC values; furthermore, the edges of the vertical bars need not correspond to the 2$\sigma$ limits of the WC in question.  The value for \tllq in the 4$\ell$ category is shown off-scale to preserve the legibility for the rest of the plot.  Despite the large ratio, the expected yield in this category is still exceptionally small after the fit and has a negligible contribution to the 2$\sigma$ confidence intervals.


\clearpage

\section{Summary}
\label{sec:summary}
A search for new physics has been performed in the production 
of at least one top quark in association with additional leptons, jets, and  \PQb jets,
in the context of an effective field theory. The events were produced in proton-proton collisions 
corresponding to an integrated luminosity of \lumi.  
The expected yield in each category was parameterized in terms of 16
Wilson coefficients (WCs) associated with effective field theory operators relevant to the
dominant processes in the data.

A simultaneous fit was performed
of the 16 WCs to the data.  For each WC, an interval over which the model predictions agree
with the observed yields at the 2 standard deviation level was extracted 
by either keeping the other WCs fixed to
zero or treating the other WCs as unconstrained nuisance parameters.  
Two-dimensional contours were produced for some of the WCs,
to illustrate correlations between various WCs.  
The results from fitting the WCs in the dimension-six model to the data were consistent 
with the standard model at the level of 2 standard deviations.

\begin{acknowledgments}
  We thank Adam Martin and Jeong Han Kim for their theoretical guidance in configuring and debugging the EFT model used to generate the signal samples in this analysis.

  We congratulate our colleagues in the CERN accelerator departments for the excellent performance of the LHC and thank the technical and administrative staffs at CERN and at other CMS institutes for their contributions to the success of the CMS effort. In addition, we gratefully acknowledge the computing  centers and personnel of the Worldwide LHC Computing Grid for delivering so effectively the computing infrastructure essential to our analyses. Finally, we acknowledge the enduring support for the construction and operation of the LHC and the CMS detector provided by the following funding agencies: BMBWF and FWF (Austria); FNRS and FWO (Belgium); CNPq, CAPES, FAPERJ, FAPERGS, and FAPESP (Brazil); MES (Bulgaria); CERN; CAS, MoST, and NSFC (China); COLCIENCIAS (Colombia); MSES and CSF (Croatia); RIF (Cyprus); SENESCYT (Ecuador); MoER, ERC PUT and ERDF (Estonia); Academy of Finland, MEC, and HIP (Finland); CEA and CNRS/IN2P3 (France); BMBF, DFG, and HGF (Germany); GSRT (Greece); NKFIA (Hungary); DAE and DST (India); IPM (Iran); SFI (Ireland); INFN (Italy); MSIP and NRF (Republic of Korea); MES (Latvia); LAS (Lithuania); MOE and UM (Malaysia); BUAP, CINVESTAV, CONACYT, LNS, SEP, and UASLP-FAI (Mexico); MOS (Montenegro); MBIE (New Zealand); PAEC (Pakistan); MSHE and NSC (Poland); FCT (Portugal); JINR (Dubna); MON, RosAtom, RAS, RFBR, and NRC KI (Russia); MESTD (Serbia); SEIDI, CPAN, PCTI, and FEDER (Spain); MOSTR (Sri Lanka); Swiss Funding Agencies (Switzerland); MST (Taipei); ThEPCenter, IPST, STAR, and NSTDA (Thailand); TUBITAK and TAEK (Turkey); NASU (Ukraine); STFC (United Kingdom); DOE and NSF (USA).
  
  \hyphenation{Rachada-pisek} Individuals have received support from the Marie-Curie program and the European Research Council and Horizon 2020 Grant, contract Nos.\ 675440, 724704, 752730, and 765710 (European Union); the Leventis Foundation; the A.P.\ Sloan Foundation; the Alexander von Humboldt Foundation; the Belgian Federal Science Policy Office; the Fonds pour la Formation \`a la Recherche dans l'Industrie et dans l'Agriculture (FRIA-Belgium); the Agentschap voor Innovatie door Wetenschap en Technologie (IWT-Belgium); the F.R.S.-FNRS and FWO (Belgium) under the ``Excellence of Science -- EOS" -- be.h project n.\ 30820817; the Beijing Municipal Science \& Technology Commission, No. Z191100007219010; the Ministry of Education, Youth and Sports (MEYS) of the Czech Republic; the Deutsche Forschungsgemeinschaft (DFG) under Germany's Excellence Strategy -- EXC 2121 ``Quantum Universe" -- 390833306; the Lend\"ulet (``Momentum") Program and the J\'anos Bolyai Research Scholarship of the Hungarian Academy of Sciences, the New National Excellence Program \'UNKP, the NKFIA research grants 123842, 123959, 124845, 124850, 125105, 128713, 128786, and 129058 (Hungary); the Council of Science and Industrial Research, India; the HOMING PLUS program of the Foundation for Polish Science, cofinanced from European Union, Regional Development Fund, the Mobility Plus program of the Ministry of Science and Higher Education, the National Science Center (Poland), contracts Harmonia 2014/14/M/ST2/00428, Opus 2014/13/B/ST2/02543, 2014/15/B/ST2/03998, and 2015/19/B/ST2/02861, Sonata-bis 2012/07/E/ST2/01406; the National Priorities Research Program by Qatar National Research Fund; the Ministry of Science and Higher Education, project no. 0723-2020-0041 (Russia); the Tomsk Polytechnic University Competitiveness Enhancement Program; the Programa Estatal de Fomento de la Investigaci{\'o}n Cient{\'i}fica y T{\'e}cnica de Excelencia Mar\'{\i}a de Maeztu, grant MDM-2015-0509 and the Programa Severo Ochoa del Principado de Asturias; the Thalis and Aristeia programs cofinanced by EU-ESF and the Greek NSRF; the Rachadapisek Sompot Fund for Postdoctoral Fellowship, Chulalongkorn University and the Chulalongkorn Academic into Its 2nd Century Project Advancement Project (Thailand); the Kavli Foundation; the Nvidia Corporation; the SuperMicro Corporation; the Welch Foundation, contract C-1845; and the Weston Havens Foundation (USA).\end{acknowledgments}

\bibliography{auto_generated}
\cleardoublepage \appendix\section{The CMS Collaboration \label{app:collab}}\begin{sloppypar}\hyphenpenalty=5000\widowpenalty=500\clubpenalty=5000\vskip\cmsinstskip
\textbf{Yerevan Physics Institute, Yerevan, Armenia}\\*[0pt]
A.M.~Sirunyan$^{\textrm{\dag}}$, A.~Tumasyan
\vskip\cmsinstskip
\textbf{Institut f\"{u}r Hochenergiephysik, Wien, Austria}\\*[0pt]
W.~Adam, T.~Bergauer, M.~Dragicevic, A.~Escalante~Del~Valle, R.~Fr\"{u}hwirth\cmsAuthorMark{1}, M.~Jeitler\cmsAuthorMark{1}, N.~Krammer, L.~Lechner, D.~Liko, I.~Mikulec, F.M.~Pitters, N.~Rad, J.~Schieck\cmsAuthorMark{1}, R.~Sch\"{o}fbeck, M.~Spanring, S.~Templ, W.~Waltenberger, C.-E.~Wulz\cmsAuthorMark{1}, M.~Zarucki
% ...
\vskip\cmsinstskip
\dag: Deceased\\
1:  Also at Vienna University of Technology, Vienna, Austria\\
% ...
\end{sloppypar}
%%% END EDITABLE REGION %%%
% skeleton_end
\end{document}